\documentclass[aps,prl,twocolumn,notitlepage,amsmath,amssymb,superscriptaddress]{revtex4-2}
\usepackage{graphicx}
\usepackage{dcolumn}
\usepackage{amsthm}
\usepackage{bm}

\newcommand{\ket}[1]{|#1\rangle}
\newcommand{\bra}[1]{\langle#1|}

\pdfoutput=1
\usepackage{graphicx,graphics,epsfig,subfigure,times,bm,bbm,amssymb,amsmath,amsthm,mathrsfs,MnSymbol}
\usepackage{gensymb}
\usepackage{amsfonts}
\usepackage[pdfstartview=FitH]{hyperref}
\hypersetup{
    colorlinks=true,       
    linkcolor=red,          
    citecolor=magenta,        
    filecolor=magenta,      
    urlcolor=cyan,           
    runcolor=cyan}
\usepackage{epstopdf}
\usepackage[pdftex]{color}
\usepackage{braket}
\usepackage{enumerate}
\usepackage[normalem]{ulem}
\usepackage[usenames,dvipsnames]{xcolor}
\usepackage{multirow}
\usepackage{mathtools}

\definecolor{orange}{rgb}{1,0.5,0}

\newcommand{\ignore}[1]{}

\usepackage{changes}
    \definechangesauthor[name={yhshi},color=blue]{per}
    \definechangesauthor[name={wyy},color=red]{wyy}
\begin{document}

\newcommand{\beginsupplement}{%
        \setcounter{table}{0}
        \renewcommand{\thetable}{S\arabic{table}}%
        \setcounter{figure}{0}
        \renewcommand{\thefigure}{S\arabic{figure}}%
        \setcounter{section}{0}
        \setcounter{equation}{0}
        \renewcommand{\theequation}{S\arabic{equation}}%
}

\title{Observation of critical phase transition in a generalized Aubry-Andr\'e-Harper model on a superconducting quantum processor with tunable couplers}

\author{Hao Li}
\altaffiliation[]{These authors contributed equally to this work.}
\affiliation{School of Physics, Northwest University, Xi'an 710127, China}
\affiliation{Institute of Physics, Chinese Academy of Sciences, Beijing 100190, China}

\author{Yong-Yi Wang}
\altaffiliation[]{These authors contributed equally to this work.}
\affiliation{Institute of Physics, Chinese Academy of Sciences, Beijing 100190, China}
\affiliation{School of Physical Sciences, University of Chinese Academy of Sciences, Beijing 100190, China}

\author{Yun-Hao Shi}
\affiliation{Institute of Physics, Chinese Academy of Sciences, Beijing 100190, China}
\affiliation{School of Physical Sciences, University of Chinese Academy of Sciences, Beijing 100190, China}

\author{Kaixuan Huang}
\affiliation{Beijing Academy of Quantum Information Sciences, and CAS Center of Excellence for Topological Quantum Computation, University of Chinese Academy of Sciences, Beijing 100190, China}
\affiliation{Institute of Physics, Chinese Academy of Sciences, Beijing 100190, China}

\author{Xiaohui Song}
\affiliation{Institute of Physics, Chinese Academy of Sciences, Beijing 100190, China}

\author{Gui-Han Liang}
\affiliation{Institute of Physics, Chinese Academy of Sciences, Beijing 100190, China}
\affiliation{School of Physical Sciences, University of Chinese Academy of Sciences, Beijing 100190, China}

\author{Zheng-Yang Mei}
\affiliation{Institute of Physics, Chinese Academy of Sciences, Beijing 100190, China}
\affiliation{School of Physical Sciences, University of Chinese Academy of Sciences, Beijing 100190, China}

\author{Bozhen Zhou}
\affiliation{Institute of Physics, Chinese Academy of Sciences, Beijing 100190, China}
\affiliation{School of Physical Sciences, University of Chinese Academy of Sciences, Beijing 100190, China}

\author{He Zhang}
\affiliation{Institute of Physics, Chinese Academy of Sciences, Beijing 100190, China}
\affiliation{School of Physical Sciences, University of Chinese Academy of Sciences, Beijing 100190, China}

\author{Jia-Chi Zhang}
\affiliation{Institute of Physics, Chinese Academy of Sciences, Beijing 100190, China}
\affiliation{School of Physical Sciences, University of Chinese Academy of Sciences, Beijing 100190, China}

\author{Shu Chen}
\affiliation{Institute of Physics, Chinese Academy of Sciences, Beijing 100190, China}

\author{Shiping Zhao}
\affiliation{Institute of Physics, Chinese Academy of Sciences, Beijing 100190, China}

\author{Ye Tian}
\affiliation{Institute of Physics, Chinese Academy of Sciences, Beijing 100190, China}

\author{Zhan-Ying Yang}
\affiliation{School of Physics, Northwest University, Xi'an 710127, China}
\affiliation{Peng Huanwu Center for Fundamental Theory, and Shaanxi Key Laboratory for Theoretical Physics Frontiers, Xi'an 710127, China}

\author{Zhongcheng Xiang}
\affiliation{Institute of Physics, Chinese Academy of Sciences, Beijing 100190, China}
\affiliation{School of Physical Sciences, University of Chinese Academy of Sciences, Beijing 100190, China}
\affiliation{Hefei National Laboratory, Hefei 230088, China}
\affiliation{Beijing Academy of Quantum Information Sciences, and CAS Center of Excellence for Topological Quantum Computation, University of Chinese Academy of Sciences, Beijing 100190, China}
\affiliation{Songshan Lake  Materials Laboratory, Dongguan 523808, Guangdong, China}

\author{Kai Xu}
\email{kaixu@iphy.ac.cn}
\affiliation{Institute of Physics, Chinese Academy of Sciences, Beijing 100190, China}
\affiliation{School of Physical Sciences, University of Chinese Academy of Sciences, Beijing 100190, China}
\affiliation{Hefei National Laboratory, Hefei 230088, China}
\affiliation{Beijing Academy of Quantum Information Sciences, and CAS Center of Excellence for Topological Quantum Computation, University of Chinese Academy of Sciences, Beijing 100190, China}
\affiliation{Songshan Lake  Materials Laboratory, Dongguan 523808, Guangdong, China}

\author{Dongning Zheng}
\email{dzheng@iphy.ac.cn}
\affiliation{Institute of Physics, Chinese Academy of Sciences, Beijing 100190, China}
\affiliation{School of Physical Sciences, University of Chinese Academy of Sciences, Beijing 100190, China}
\affiliation{Hefei National Laboratory, Hefei 230088, China}
\affiliation{Songshan Lake  Materials Laboratory, Dongguan 523808, Guangdong, China}

\author{Heng Fan}
\email{hfan@iphy.ac.cn}
\affiliation{Institute of Physics, Chinese Academy of Sciences, Beijing 100190, China}
\affiliation{School of Physical Sciences, University of Chinese Academy of Sciences, Beijing 100190, China}
\affiliation{Hefei National Laboratory, Hefei 230088, China}
\affiliation{Beijing Academy of Quantum Information Sciences, and CAS Center of Excellence for Topological Quantum Computation, University of Chinese Academy of Sciences, Beijing 100190, China}
\affiliation{Songshan Lake  Materials Laboratory, Dongguan 523808, Guangdong, China}

\begin{abstract}
Quantum simulation enables study of many-body systems in non-equilibrium by mapping to a controllable quantum system, providing a new tool for computational intractable problems. Here, using a programmable quantum processor with a chain of 10 superconducting qubits interacted through tunable couplers, we simulate the one-dimensional generalized Aubry-Andr\'e-Harper model for three different phases, i.e., extended, localized and critical phases. The properties of phase transitions and many-body dynamics are studied in the presence of quasi-periodic modulations for both off-diagonal hopping coefficients and on-site potentials of the model controlled respectively by adjusting strength of couplings and qubit frequencies. 
We observe the spin transport for initial single- and multi-excitation states in different phases, and characterize phase transitions by experimentally measuring dynamics of participation entropies. Our experimental results demonstrate that the newly developed tunable coupling architecture of 
superconducting processor extends greatly the simulation realms for a wide variety of Hamiltonians, and may trigger further investigations on various quantum and topological phenomena.
\end{abstract}

\pacs{Valid PACS appear here}
\maketitle

\textit{Introduction.}---%
Using controllable quantum systems, quantum simulation provides a powerful approach to study
many-body physics, which might be challenging for a classical computer~\cite{Feynman1982,Georgescu2014}. In analogue quantum simulation, specific model Hamiltonians can be directly realized by engineering the platform Hamiltonians such that
dynamics of real quantum systems can be studied in a controllable manner, such as in trapped ions~\cite{Blatt2012,Zhang2017,Morong2021}, atoms in optical lattices~\cite{Bloch2012,Gross2017,Saffman2016,Bernien2017}, superconducting qubits~\cite{Makhlin2001,Xiang2013,Gu2017}, and nuclear spins~\cite{vandersypen2005nmr,Rovny2018}. 
Particularly, superconducting quantum simulation can explore a wide regime from localization to weak and strong thermalization in non-equilibrium quantum many-body systems \cite{Chen2021,Roushan2017,Xu2018,Guo2020,Guo2021,Mi2022,Yan2019,Gong2021,Braumuller2022}.

On the other hand, the 1D Aubry-Andr\'e-Harper (AAH) model~\cite{Harper1955,aubry1980}, as a workhorse for studying localization and topological states, has attracted much attention both theoretically and experimentally~\cite{Thouless1983,Ostlund1983,Hiramoto1989,Roati2008,Lahini2009,Kraus2012,Schreiber2015,Bordia2017,Roushan2017}. 
The original AAH model can be derived from a 2D quantum Hall system with nearest-neighbor hopping. When considering the next-nearest-neighbor hopping, one can deduce a generalization of the AAH model with both on-site and off-diagonal quasi-periodic modulations~\cite{Hatsugai1990,Han1994}. The generalized AAH (GAAH) model shows different and interesting localization and topological properties, for instance, the critical phase featured by multifractal wave functions and the topological adiabatic pumping~\cite{Chang1997,Takada2004,Gong2008,Liu2015,Zhao2017,Wang2021}. With the development of experimental technologies, the GAAH model has been realized in photonic crystals with on-site or off-diagonal modulation~\cite{Kraus2012} and cold atoms systems in momentum space~\cite{Xiao2021}. With flexible control and precise measurement of superconducting processor, the GAAH model may be simulated analogously, given off-diagonal quasi-periodic modulations can be implemented precisely.

\begin{figure*}[t]
	\centering
	\includegraphics[width=0.85\linewidth]{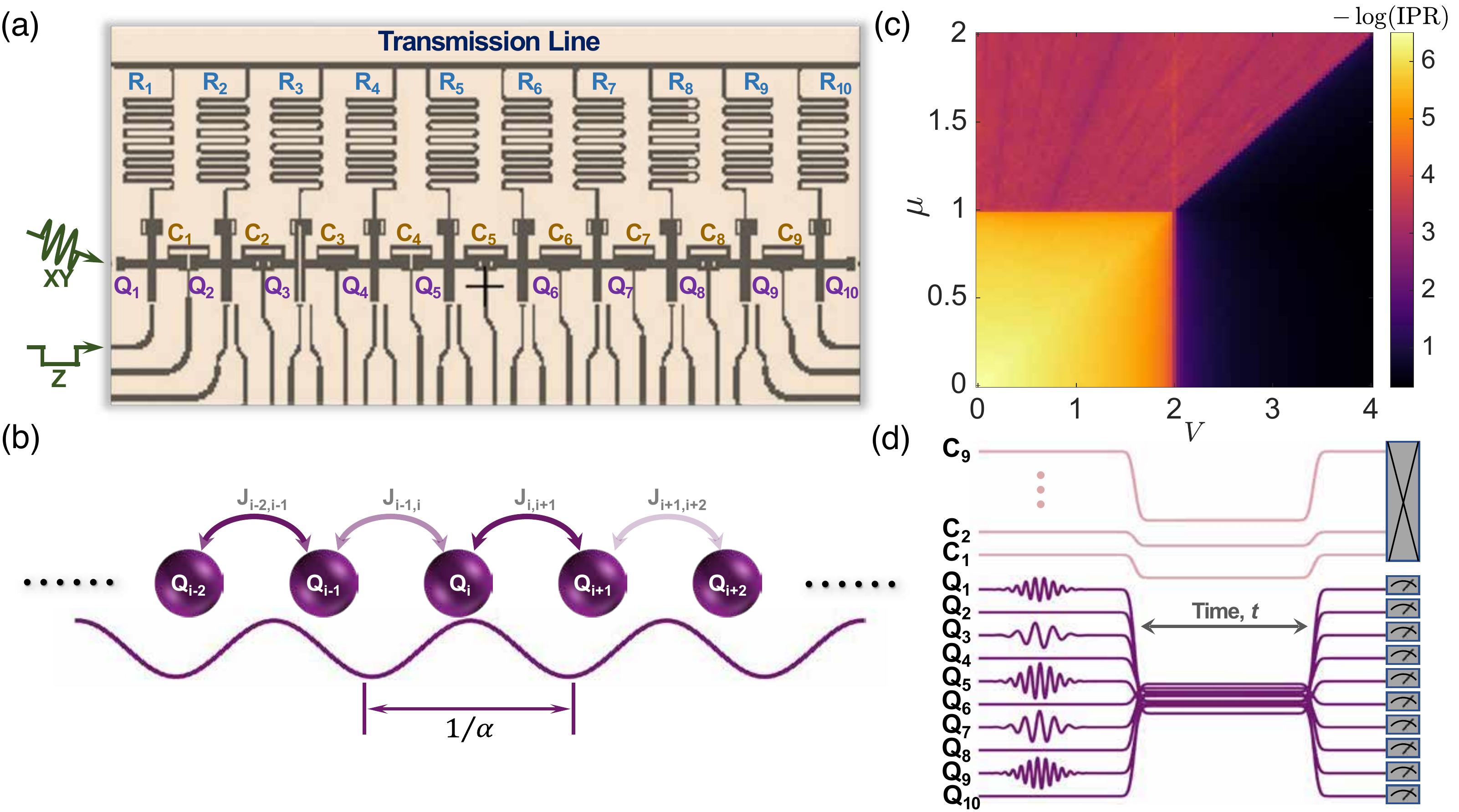}\\
	\caption{(a) Circuit diagram of the superconducting processor, consisting of ten transmon qubits, $Q_1$--$Q_{10}$, and nine couplers $C_1$--$C_9$, each of which is placed between every two nearest-neighbor qubits. (b) Schematic representation of the GAAH model with both the on-site and off-diagonal quasi-periodic modulations, with $\alpha=\frac{\sqrt{5}-1}{2}$ throughout the work. (c) Phase diagram of the GAAH model, divided into extended, localized, and critical phases. The heatmap shows the inverse participation ratio (IPR) averaged over all eigenstates and 100 random global phase offset $\delta$ for system size $L=1000$. (d) Experimental pulse sequences for observing localization properties of the GAAH model.}
    \label{fig1}
\end{figure*}

In our experiment, taking advantage of newly developed tunable coupling architecture~\cite{Yan2018,Shi2021}, we simulate the GAAH model for a wide variety of parameters on a superconducting processor. By adjusting both qubits and couplers, we experimentally observe the dynamics of the extended, localized, and critical phase in the GAAH model, and investigate the phase transition from the perspective of non-equilibrium dynamics. We observe that in the critical phase, the spin can propagate over a range intermediate between that of the extended phase and of the localized phase, for both initial single- and multi-excitation states. In addition, we quantify how fast initial states spread over the Hilbert space for different phases by experimentally measuring the time evolution of participation entropies, and characterize the transition among the extended, localized, and critical phase by calculating averaged late-time participation entropies. 

\begin{figure*}[t]
	\centering
	\includegraphics[width=1\linewidth]{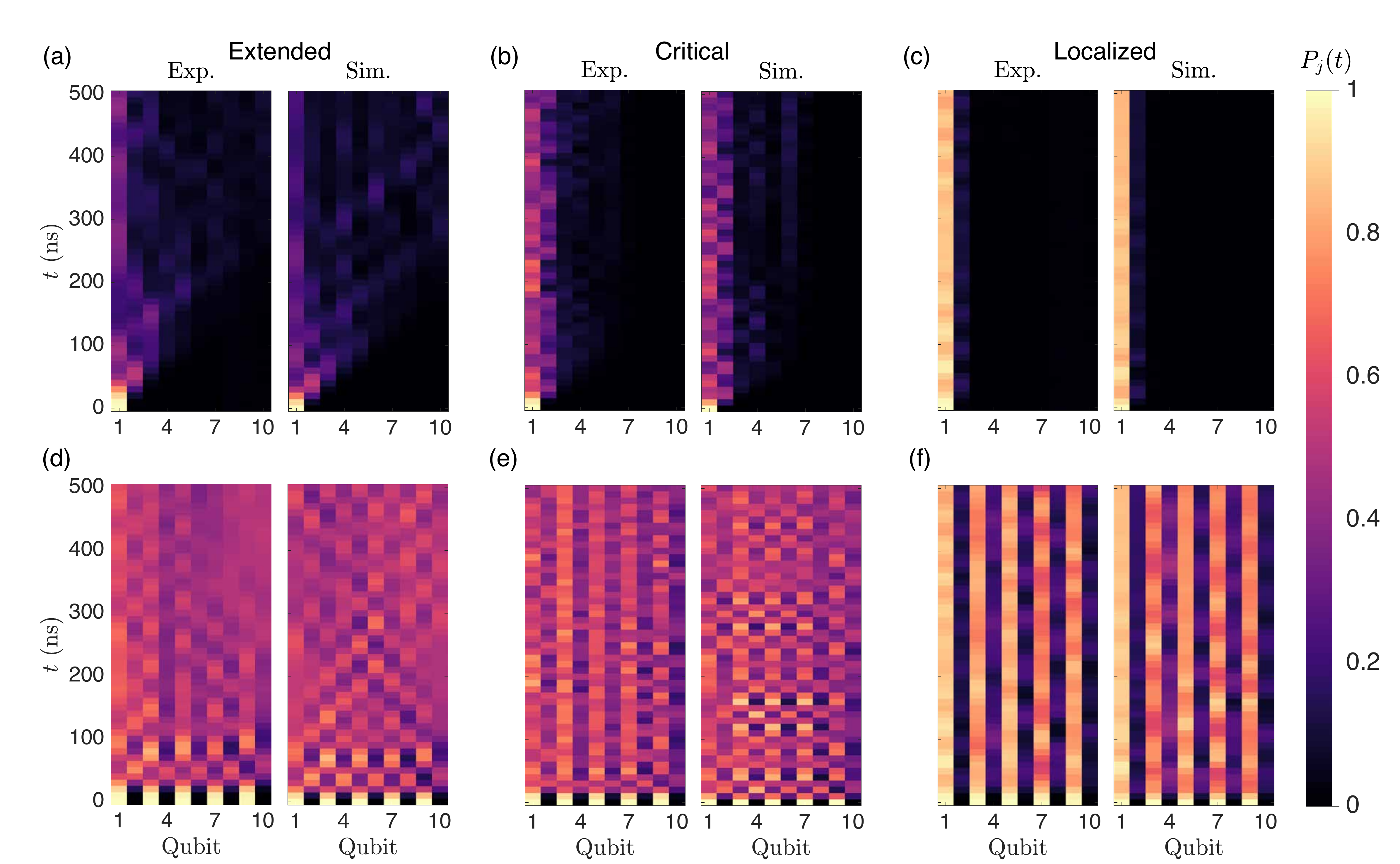}\\
	\caption{The time evolutions of qubit-resolved photon occupancy probabilities $P_j(t)$ for the system initialized in (a--c) $\ket{\psi(0)}=\ket{1000000000}$ , and (d--f) $\ket{\psi(0)}=\ket{1010101010}$, in (a, d) the extended phases with $\mu=0.5$ and $V=0.5$; (b, e) the critical phases with $\mu=2.0$ and $V=0.5$; (c, f) the localized phases with $\mu=0.5$ and $V=4.0$. The left panel of each figure shows experimental data, and the right panel shows numerical simulation. Experimental data are averaged over 5 realizations of random chosen phases $\delta$, while numerical simulation are averaged over 50 realizations of random chosen phases $\delta$ with decoherence taken into account.} 
    \label{fig2}
\end{figure*}

\textit{Model and set-up.}---%
Our quantum processor consists of a chain array of $L = 10$ transmon superconducting qubits, and 9 tunable couplers with each placed between every two nearest-neighbor qubits, which enable an accurate control of couplings (see Fig.~\ref{fig1}(a) and Supplementary Material for details). The effective Hamiltonian of the qubits system can be described by the Bose-Hubbard model,
\begin{equation} 
	\frac{\hat H}{\hbar} = \sum\limits_{ j =1 } ^{9}{(J_{j,j+1} {\hat a}_j^\dagger {\hat a}_{j+1} + \text{H.c.})}+ \sum\limits_{ j =1 } ^{10}{h_j} {\hat n_j}-\frac{U_j}{2}{{\hat n}_j} ({\hat n}_j-1),
    \end{equation}
where ${\hat a}_j^\dagger (\hat a_j)$ is the photon creation (annihilation) operator, ${\hat n}_j\equiv{\hat a}_j^\dagger \hat a_j$ is the number operator, $J_{j,j+1}$ is the tunable nearest-neighbor coupling strength, $h_{j}$ is the tunable local potential, and $U_{j}$ denotes the qubit anharmonicity serving as the on-site interaction. Since $U_j>>J_{i,j}$ for our processor, an excess of energy is needed for having more than one photon at each site, so the system can be described by hard-core bosons, or equivalently spin-$1/2$ spins, with the conservation of the total photons (or spins) due to the $U(1)$ symmetry.

In our superconducting processor with tunable couplers, the coupling $J_{j,j+1}$ contains two part: (\expandafter{\romannumeral1}) direct coupling between nearest-neighbor qubits $J_{j,j+1}^{0}$ and (\expandafter{\romannumeral2}) superexchange interaction via the coupler in between $J_{j,j+1}^{SE} \propto 1/\Delta_{j,j+1}$ with $\Delta_{j,j+1}$ denoting frequency detuning between the $j$-th coupler and the two nearest-neighbor qubits (see Supplementary Material for details). Therefore, by applying fast voltages to the Z control lines of the couplers, $J_{j,j+1}$ can be tuned individually from $-30$ to $+4.8$ MHz in our processor. In addition, by applying the fast Z pulse on each qubit, the local potential can also be arbitrarily tuned relative to resonant frequency $\sim4.36$ GHz. In this way, we can finally realize a hard-core bosonic version of the GAAH model,
\begin{eqnarray} \nonumber
    \frac{\hat H }{\hbar}= &&\lambda\sum\limits_{j=1}^{9} {\left( 1 + \mu \cos \left[ {2\pi \left( {j + \frac{1}{2}} \right)\alpha  + \delta } \right]\right) {\hat a}_j^{\dagger} {{\hat a}_{j + 1}}+ \text{H.c.}}\\
    &&+\lambda \sum\limits_{j=1}^{10} {V\cos (2\pi j\alpha  + \delta ) {\hat a}_j^{\dagger} {{\hat a}_{j}}}.
    \label{eq_GAAH}
    \end{eqnarray}
Here, as depicted in Fig.~\ref{fig1}(b), we take a quasi-periodic modulation $J_{j, j+1} = \lambda \left( 1 + \mu \cos \left[ {2\pi \left( {j + 1/2} \right)\alpha  + \delta } \right]\right)$ and $h_j = \lambda V\cos (2\pi j\alpha  + \delta )$, where $\mu$ and $V$ indicate the off-diagonal and on-site modulations amplitudes, respectively, and we choose $\lambda/(2\pi)=4$ MHz throughout the work. Besides, $\alpha=(\sqrt{5}-1)/2$ is the irrational frequency which takes the same value for on-site and off-diagonal modulations, and $\delta \in \left[-\pi , \pi\right)$ is an arbitrary global phase offset.  

For different parameters $\mu$ and $V$, localization property of the eigenstates of Hamiltonian~Eq.\eqref{eq_GAAH} can be characterized by the inverse participation ratio (IPR)~\cite{THOULESS197493}. In Fig.~\ref{fig1}(c), we plot the localization phase diagram of Hamiltonian~Eq.\eqref{eq_GAAH}. The heatmap shows the eigenstate's IPR $=\sum_{i}\left|\psi_{n,i}\right|^{4}$, where $\psi_{n,i} = \langle \psi_n| i\rangle$ is the wave function coefficient of the eigenstate $\ket{\psi_n}$ expressed in the computational basis $\{\ket{i}\}$. Here, we actually displays the negative logarithm of the IPR, in order to distinguish the three phases more clearly, and to associate with the participation entropy in the next section. Since mobility edges are absent in our model~\cite{Liu2015}, (\romannumeral1) for $V < 2, \mu < 1$, all bulk eigenstates are extended, denoted as the extended phase, where the IPR vanishes and its negative logarithm is close to $\log{L} \sim 6.9$; (\romannumeral2) for $V > 2\max(1,\mu)$, all bulk eigenstates are localized with the IPR close to one and its negative logarithm close to zero, denoted as the localized phase; (\romannumeral3) in the rest, the eigenstates are critical with intermediate IPRs, denoted as the critical phase.

\begin{figure*}[t]
	\centering
	\includegraphics[width=1\linewidth]{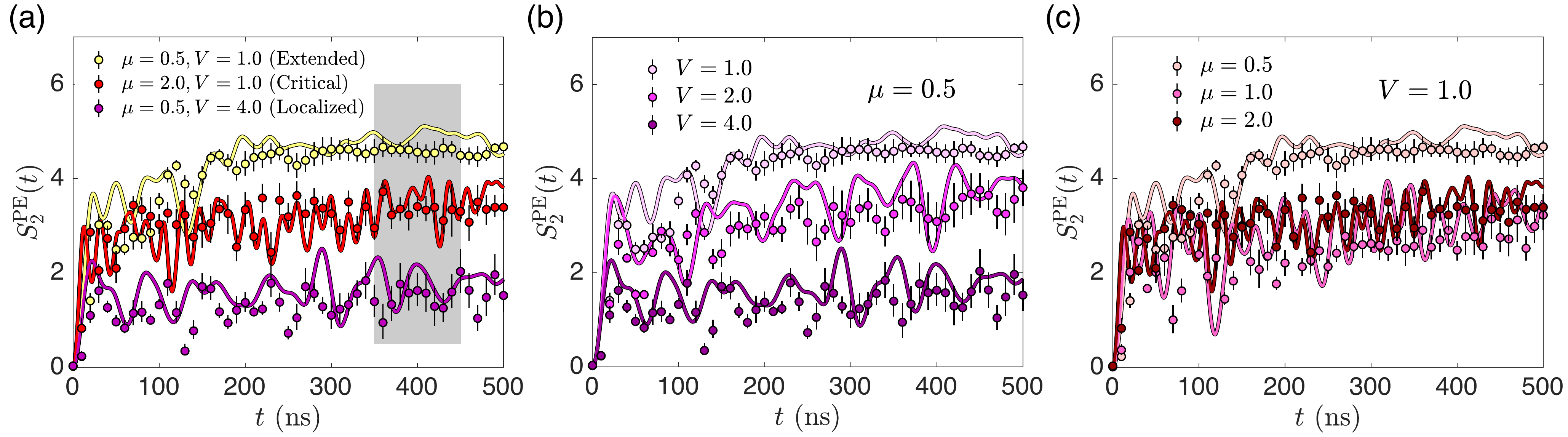}\\
	\caption{(a) The dynamics of participation entropy for the system quenched into the three phases. The yellow, red, and purple points are experimental data for the extended ($\mu=0.5, V=1.0$), critical ($\mu=2.0, V=1.0$), and localized ($\mu=0.5, V=4.0$) phases, respectively. Lines with the same color are numerical simulation under the same parameters with decoherence taken into account. (b) The dynamics of participation entropy along the path \uppercase\expandafter{\romannumeral1} in Fig.~\ref{fig4}(a) (extended to localized transition)  with fixed $\mu=0.5$ and varying $V=1$ to $4$. (c) The dynamics of participation entropy along the path \uppercase\expandafter{\romannumeral2} (extended to critical transition) with fixed $V=1$ and varying $\mu=0.5$ to $2$. Averages are taken among 10 initial states, including $\ket{\psi_{1\text{--}5}(0)}=\ket{1010101010}, \ket{1010101001}, \ket{1010100101}, \ket{1010010101}, \ket{1001010101}$, and $\ket{\psi_{6\text{--}10}(0)}=\hat{G} \ket{\psi_{1\text{--}5}(0)} $ with a global spin-flip operator $\hat{G}=\prod \limits_{j=1}^{L}\hat{\sigma}^x_j$.}
    \label{fig3}
\end{figure*}

\textit{Observation of spin transport in the GAAH model.}---%
First, we show that the GAAH model can be simulated with our superconducting qubits in a tunable coupling architecture. The experimental scheme is that three phases of the GAAH model manifest distinguishing localization properties by measuring the photon occupancy probabilities for specific initial states. We initially excite the leftmost qubit $Q_1$, i.e., the system is initialized as $\ket{\psi(0)}=\ket{1000000000}$, where $\ket{0}$ ($\ket{1}$) denotes the ground (excited) state of a qubit. Then we apply the fast Z pulse on each qubit and coupler, and the system will evolve under the Hamiltonian~Eq.\eqref{eq_GAAH}, satisfying Schr\"{o}dinger equation $\ket{\psi(t)}=e^{-i\hat{H}t}\ket{\psi(0)}$. We monitor its dynamics from $t=0$ to $500$ ns, by measuring the photon occupancy probabilities of each qubit $P_{j}(t)=\left\langle\psi(t)\left|\hat a_{j}^{\dagger} \hat a_{j}\right| \psi(t)\right\rangle$. For each time point, we perform 5000 repeated single-shot measurements.

The experimental results for the three phases are plotted in the left panel of Fig.~\ref{fig2}(a-c), with a comparison of numerical simulations in the right panel of Fig.~\ref{fig2}(a-c). As shown in Fig.~\ref{fig2}(a), in the extended region, the spin transport is not blocked, and a lightcone-like propagation and reflection at the boundary are still visible when weak off-diagonal and on-site quasi-periodic disorder  exists. As the opposite, for sufficiently large on-site disorder $V$, the spin is fully localized, and only the initially occupied site has a occupancy probability close to one at any time (Fig.~\ref{fig2}(c)). In the critical region, the spin tends to oscillate around adjacent sites of the initially occupied site, and the propagation range is intermediate between the above two (Fig.~\ref{fig2}(b)).

With the capability of precise simultaneous control and readout, we can also prepare initial product states to probe the largest Hilbert space (i.e., the half-filled sector), by exciting selected $M=N/2=5$ qubits and keeping the rest in their ground state. The experimental sequences are shown in Fig.~\ref{fig1}(d). Here, we focus on an initial N\'eel state $\ket{\psi(0)}=\ket{1010101010}$, and measure the photon occupancy probabilities from $t=0$ to $500$ ns. The experimental and numerical results are plotted in Fig.~\ref{fig2}(d-f). As shown in Fig.~\ref{fig2}(d), in the extended region, the mean photon occupancy probabilities oscillates around $0.5$ at long times for all ten qubits with a small fluctuation, showing a pattern of oscillation between odd and even sites back and forth. In Fig.~\ref{fig2}(e), the photon occupancy probabilities for all qubits are also close to 0.5 in the critical region. However, different from the extended region, the photon occupancy probabilities in the critical region exhibit a certain degree of dependence on the initial configuration, that is, the photon occupancy probabilities at the initially occupied sites tend to stay above $0.5$, while those at initially unoccupied sites tend to stay below $0.5$. Figure~\ref{fig2}(f) shows the spin transport is completely blocked in the localized region, and the occupancy probabilities stay close to one for the initially occupied sites (the odd sites), and close to zero for initially unoccupied sites (the even sites) at any time.

\textit{Dynamical signature of localization via participation entropies.}---%
As stated above, localization properties can be quantified by the IPR and the associated participation entropy, which are used to describe single particle localization~\cite{Bell1972,Schfer1980,Rodriguez2011} and is recently used in many-body physics~\cite{Luitz2014,Luitz2015}. Here, we define the $q$-th order dynamical participation entropy as 
\begin{equation}
S^{\text{PE}}_q(t) = \frac{1}{1-q}\log{\sum\limits_{i}^{\mathcal{N}} p_{i}(t)^q},
\label{PE}
\end{equation}
where $\mathcal{N}$ is the dimension of Hilbert space, and $p_{i}(t) = |\langle \psi_t| i\rangle|^{2}$. Considering the $U(1)$ symmetry, the dimension of the half-filled sector is $\mathcal{N}= {10 \choose 5} = 252$. Here, we focus on the second-order participation entropy, i.e., $S^{\text{PE}}_2(t) = -\log{\sum\limits_{i}^{\mathcal{N}} p_{i}(t)^2}$, which is related to IPR by taking the negative logarithm. In the Supplementary Material, we also display the results of first-order participation entropy.

The dynamical participation entropy is a characterization quantifying how fast $\ket{\psi(t)}$ spreads over the Hilbert space. Here we select $2\times M=10$ initial states which is far from equilibrium, taking the form $\ket{\psi_i(0)}=\ket{10}^{\otimes(M+1-i)}\bigotimes\ket{01}^{\otimes(i-1)}$, and $\ket{\psi_{M+i}(0)}=\hat{G} \ket{\psi_{i}(0)}$ with a global spin-flip operator $\hat{G}=\prod \limits_{j=1}^L\hat{\sigma}^x_j$, for $i=1,\ldots,M$. The experimental data shown in Fig~\ref{fig3} are averaged over these 10 initial states, and the measured multi-qubit probabilities $p_{i}$ are post-selected within the half-filled sector due to the $U(1)$ symmetry before calculating the participation entropy.

Figure~\ref{fig3}(a) displays the time evolution of participation entropy for the system quenched into the three phases. After a fast initial relaxation, the participation entropy oscillates around some certain value, which varies in different phases. For both small $\mu$ and $V$, the system lies in the extended phase, where the late-time participation entropy keeps at a high value with a small oscillation (see the yellow points in Fig.~\ref{fig3}(a)). For small $\mu$ and sufficiently large $V$, the participation entropy oscillates around a much smaller value (see the purple points), as we would expect in the localized phase. For relatively large $\mu$, the late-time participation entropy stands between the extended and localized cases (see the purple points), which behaves similarly to that with a small $\mu$ and intermediate $V$ around the extended to localized transition point (see the data of $V=2$ in Fig.~\ref{fig3}(b) for comparison).

Figure~\ref{fig3}(b) and (c) show the time evolution of participation entropy with increasing $V$ for fixed $\mu=0.5$, and increasing $\mu$ for fixed $V=1$, respectively. As shown in Fig.~\ref{fig3}(b), with the increase of $V$, the growth of participation entropy is suppressed significantly, reflecting a transition from the extended region to the localized region. In contrast, as $\mu$ increases to approach the theoretical transition point $\mu_c=1.0$, the growth of participation entropy slows down compared to the extended phase. However, as $\mu$ continues to increase, the curve of time evolution of participation entropy rises slightly again.
The late-time participation entropy in the critical phase reflect a multifractal behavior, and the multifractal analysis can be found in the Supplementary Material.

\begin{figure}[t]
	\centering
	\includegraphics[width=1\linewidth]{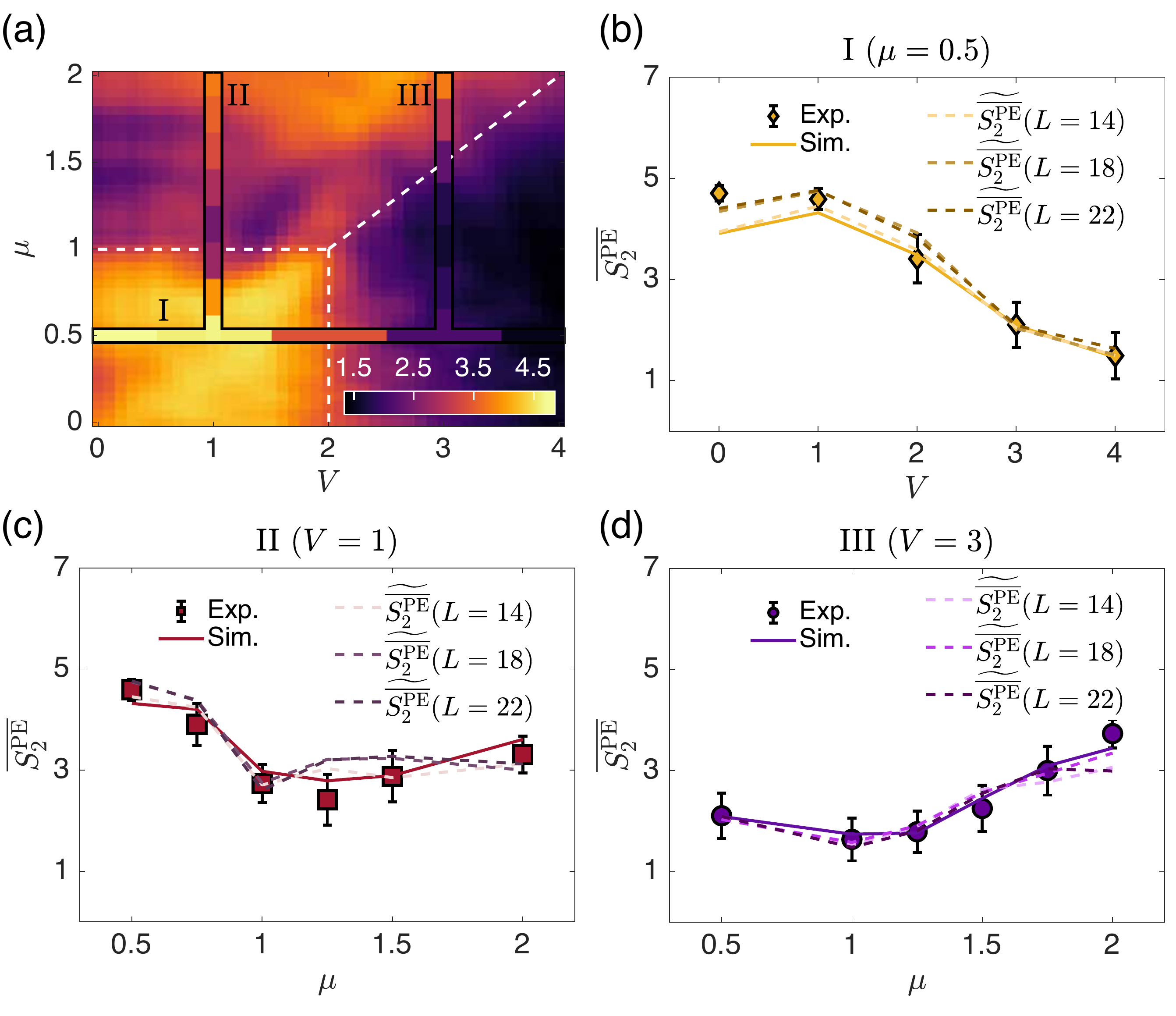}\\
	\caption{(a) Averaged late-time participation entropy $\overline{S^{\text{PE}}_2}$ as a function of $\mu$ and $V$. The stripes \uppercase\expandafter{\romannumeral1}, \uppercase\expandafter{\romannumeral2} and \uppercase\expandafter{\romannumeral3} show the experimentally measured averaged late-time participation entropy, and the underlying phase diagram shows the numerically calculated averaged late-time participation entropy using the Hamiltonian~(\ref{eq_GAAH}). The averaged late-time participation entropy $\overline{S^{\text{PE}}_2}$ is averaged over a time window $t=350$ to $450$ ns (see the grey area in Fig.~\ref{fig3}(a)). The white dashed line shows theoretical phase boundaries as a guide to eye. (b--d) Comparisons between experimental data and numerical simulation along the path (b) \uppercase\expandafter{\romannumeral1} with fixed $\mu=0.5$, (c) \uppercase\expandafter{\romannumeral2} with fixed $V=1$, and (d)  \uppercase\expandafter{\romannumeral3} with fixed $V=3$. Points with statistical error bars are experimental data, and solid lines are numerical simulation using the Hamiltonian~(\ref{eq_GAAH}).  Dashed lines exhibit the numerically calculated averaged late-time participation entropy for larger system sizes $L=14, 18$ and $22$, rescaled as $\widetilde{\overline{S^{{\text{PE}}}_2}}(L)=\frac{\log{\mathcal{N}_{10}}}{\log{\mathcal{N}_{L}}}\cdot\overline{S^{{\text{PE}}}_2}(L)$ with $\mathcal{N}_{L}={ L \choose L/2 }$ for a direct comparison to the experimental results for $L=10$.}
    \label{fig4}
\end{figure}

In addition, the averaged participation entropy at long times can be used as an experimentally accessible characterization of phase transition. Here, we experimentally measure the averaged late-time participation entropies $\overline{S^{\text{PE}}_2}$ along three paths \uppercase\expandafter{\romannumeral1}, \uppercase\expandafter{\romannumeral2} and \uppercase\expandafter{\romannumeral3} in the $\mu-V$ plane (see Fig.~\ref{fig4}(a)), corresponding to extended to localized transition, extended to critical transition, and localized to critical transition, respectively. Averages are taken over a time window from $350$ to $450$ ns (see the grey area in Fig.~\ref{fig3}(a)), and among the 10 initial states as defined before. In Fig.~\ref{fig4}(a), we also display the numerical results of the averaged late-time participation entropy calculated using Hamiltonian~(\ref{eq_GAAH}) in the same way as the experimental processing, for the whole $\mu-V$ plane ($0 \leq V \leq 4, 0 \leq \mu \leq 2$) as a reference, which exhibits a similar phase diagram to IPR averaged over eigenstates in Fig.~\ref{fig1}(c). Because the time window $350$--$450$ ns is far less than the averaged $\overline{T_1}\sim22.3~\mu$s (see Supplementary Material), and $T_2$ has little effect on the averaged late-time participation entropies, we ignore the effect of decoherence in the simulation here. The numerical simulation and experimental results are consistent well with each other.

Comparisons of the specific experimental data with numerical simulation of the three paths \uppercase\expandafter{\romannumeral1}, \uppercase\expandafter{\romannumeral2} and \uppercase\expandafter{\romannumeral3} are plotted in Fig.~\ref{fig4}(b), (c) and (d), respectively. Different from the path \uppercase\expandafter{\romannumeral1}, where $\overline{S^{\text{PE}}_2}$ decreases monotonically as $\mu$ increases, for the path \uppercase\expandafter{\romannumeral2}, $\overline{S^{\text{PE}}_2}$ first decreases to the minimum around the theoretical transition point $\mu_c=1.0$, and then increases slightly but keeps almost unchanged for increasing $\mu$ in the critical phase. Note that both the experimentally measured and numerically calculated minimum $\overline{S^{\text{PE}}_2}$ are reached at $\mu=1.25$ instead of $\mu_c=1.0$ for $L=10$, which we attribute to finite size effects. To see this, we also calculate the averaged late participation entropies for larger system sizes $L=14,\ 18,\ 22$, which we rescale by a factor $\frac{\log{\mathcal{N}_{10}}}{\log{\mathcal{N}_{L}}}$ with $\mathcal{N}_{L}={ L \choose L/2 }$, i.e., $\widetilde{\overline{S^{{\text{PE}}}_2}}(L)=\frac{\log{\mathcal{N}_{10}}}{\log{\mathcal{N}_{L}}}\cdot\overline{S^{{\text{PE}}}_2}(L)$, for a direct comparison to the experimental system size $L=10$. In the systems with $L=14,\ 18,\ 22$, the minimum $\overline{S^{\text{PE}}_2}$ are reached at $\mu_c=1.0$. For the path \uppercase\expandafter{\romannumeral3}, i.e., the localized to critical transition, $\overline{S^{\text{PE}}_2}$ decreases slightly for increasing $\mu \leq 1$, and then increases and finally reaches a intermediate value featuring a multifractal behavior, as $\mu$ increases across the transition point $\mu_c^\prime=1.5$ for fixed $V=3$.
It is worth pointing out that the slope at the transition point $\mu_c^\prime=1.5$ increases with system size, and will diverge in the thermodynamic limit, as a signature of the localized to critical transition.

\textit{Conclusions.}---%
We implement a simulation of the GAAH model for a wide range of parameters. The capability of individual control and multi-qubit simultaneous readout of our superconducting processor allows for observation of multi-qubit spin transport and measurement of participation entropies in the experiments revealing clearly phase transitions of the model. Our experiment paves the way to quantum simulation of rich many-body phases and
may trigger further studies about dynamics of many-body systems.

\begin{acknowledgments}
We thank Zheng-Hang Sun and Shang-Shu Li for valuable discussions. This work was supported by National Natural Science Foundation of China (Grants Nos. 11934018, T2121001, 11904393, 92065114, 11875220 and 12047502),
Innovation Program for Quantum Science and Technology (Grant No. 2-6),
Strategic Priority Research Program of Chinese Academy of Sciences (Grant No. XDB28000000),
Beijing Natural Science Foundation (Grant No. Z200009), and Scientific Instrument Developing Project of Chinese Academy of Sciences (Grant No. YJKYYQ20200041).
\end{acknowledgments}

\bibliography{manuscript}

\begin{thebibliography}{50}%
\makeatletter
\providecommand \@ifxundefined [1]{%
 \@ifx{#1\undefined}
}%
\providecommand \@ifnum [1]{%
 \ifnum #1\expandafter \@firstoftwo
 \else \expandafter \@secondoftwo
 \fi
}%
\providecommand \@ifx [1]{%
 \ifx #1\expandafter \@firstoftwo
 \else \expandafter \@secondoftwo
 \fi
}%
\providecommand \natexlab [1]{#1}%
\providecommand \enquote  [1]{``#1''}%
\providecommand \bibnamefont  [1]{#1}%
\providecommand \bibfnamefont [1]{#1}%
\providecommand \citenamefont [1]{#1}%
\providecommand \href@noop [0]{\@secondoftwo}%
\providecommand \href [0]{\begingroup \@sanitize@url \@href}%
\providecommand \@href[1]{\@@startlink{#1}\@@href}%
\providecommand \@@href[1]{\endgroup#1\@@endlink}%
\providecommand \@sanitize@url [0]{\catcode `\\12\catcode `\$12\catcode
  `\&12\catcode `\#12\catcode `\^12\catcode `\_12\catcode `\%12\relax}%
\providecommand \@@startlink[1]{}%
\providecommand \@@endlink[0]{}%
\providecommand \url  [0]{\begingroup\@sanitize@url \@url }%
\providecommand \@url [1]{\endgroup\@href {#1}{\urlprefix }}%
\providecommand \urlprefix  [0]{URL }%
\providecommand \Eprint [0]{\href }%
\providecommand \doibase [0]{https://doi.org/}%
\providecommand \selectlanguage [0]{\@gobble}%
\providecommand \bibinfo  [0]{\@secondoftwo}%
\providecommand \bibfield  [0]{\@secondoftwo}%
\providecommand \translation [1]{[#1]}%
\providecommand \BibitemOpen [0]{}%
\providecommand \bibitemStop [0]{}%
\providecommand \bibitemNoStop [0]{.\EOS\space}%
\providecommand \EOS [0]{\spacefactor3000\relax}%
\providecommand \BibitemShut  [1]{\csname bibitem#1\endcsname}%
\let\auto@bib@innerbib\@empty
\bibitem [{\citenamefont {Feynman}(1982)}]{Feynman1982}%
  \BibitemOpen
  \bibfield  {author} {\bibinfo {author} {\bibfnamefont {R.~P.}\ \bibnamefont
  {Feynman}},\ }\bibfield  {title} {\bibinfo {title} {{Simulating physics with
  computers}},\ }\href {https://doi.org/10.1007/BF02650179} {\bibfield
  {journal} {\bibinfo  {journal} {Int. J. Theor. Phys.}\ }\textbf {\bibinfo
  {volume} {21}},\ \bibinfo {pages} {467} (\bibinfo {year} {1982})}\BibitemShut
  {NoStop}%
\bibitem [{\citenamefont {Georgescu}\ \emph {et~al.}(2014)\citenamefont
  {Georgescu}, \citenamefont {Ashhab},\ and\ \citenamefont
  {Nori}}]{Georgescu2014}%
  \BibitemOpen
  \bibfield  {author} {\bibinfo {author} {\bibfnamefont {I.~M.}\ \bibnamefont
  {Georgescu}}, \bibinfo {author} {\bibfnamefont {S.}~\bibnamefont {Ashhab}},\
  and\ \bibinfo {author} {\bibfnamefont {F.}~\bibnamefont {Nori}},\ }\bibfield
  {title} {\bibinfo {title} {{Quantum simulation}},\ }\href
  {https://doi.org/10.1103/RevModPhys.86.153} {\bibfield  {journal} {\bibinfo
  {journal} {Rev. Mod. Phys.}\ }\textbf {\bibinfo {volume} {86}},\ \bibinfo
  {pages} {153} (\bibinfo {year} {2014})}\BibitemShut {NoStop}%
\bibitem [{\citenamefont {Blatt}\ and\ \citenamefont {Roos}(2012)}]{Blatt2012}%
  \BibitemOpen
  \bibfield  {author} {\bibinfo {author} {\bibfnamefont {R.}~\bibnamefont
  {Blatt}}\ and\ \bibinfo {author} {\bibfnamefont {C.~F.}\ \bibnamefont
  {Roos}},\ }\bibfield  {title} {\bibinfo {title} {{Quantum simulations with
  trapped ions}},\ }\href {https://doi.org/10.1038/nphys2252} {\bibfield
  {journal} {\bibinfo  {journal} {Nat. Phys.}\ }\textbf {\bibinfo {volume}
  {8}},\ \bibinfo {pages} {277} (\bibinfo {year} {2012})}\BibitemShut {NoStop}%
\bibitem [{\citenamefont {Zhang}\ \emph {et~al.}(2017)\citenamefont {Zhang},
  \citenamefont {Pagano}, \citenamefont {Hess}, \citenamefont {Kyprianidis},
  \citenamefont {Becker}, \citenamefont {Kaplan}, \citenamefont {Gorshkov},
  \citenamefont {Gong},\ and\ \citenamefont {Monroe}}]{Zhang2017}%
  \BibitemOpen
  \bibfield  {author} {\bibinfo {author} {\bibfnamefont {J.}~\bibnamefont
  {Zhang}}, \bibinfo {author} {\bibfnamefont {G.}~\bibnamefont {Pagano}},
  \bibinfo {author} {\bibfnamefont {P.~W.}\ \bibnamefont {Hess}}, \bibinfo
  {author} {\bibfnamefont {A.}~\bibnamefont {Kyprianidis}}, \bibinfo {author}
  {\bibfnamefont {P.}~\bibnamefont {Becker}}, \bibinfo {author} {\bibfnamefont
  {H.}~\bibnamefont {Kaplan}}, \bibinfo {author} {\bibfnamefont {A.~V.}\
  \bibnamefont {Gorshkov}}, \bibinfo {author} {\bibfnamefont {Z.-X.}\
  \bibnamefont {Gong}},\ and\ \bibinfo {author} {\bibfnamefont
  {C.}~\bibnamefont {Monroe}},\ }\bibfield  {title} {\bibinfo {title}
  {{Observation of a many-body dynamical phase transition with a 53-qubit
  quantum simulator}},\ }\href {https://doi.org/10.1038/nature24654} {\bibfield
   {journal} {\bibinfo  {journal} {Nature}\ }\textbf {\bibinfo {volume}
  {551}},\ \bibinfo {pages} {601} (\bibinfo {year} {2017})}\BibitemShut
  {NoStop}%
\bibitem [{\citenamefont {Morong}\ \emph {et~al.}(2021)\citenamefont {Morong},
  \citenamefont {Liu}, \citenamefont {Becker}, \citenamefont {Collins},
  \citenamefont {Feng}, \citenamefont {Kyprianidis}, \citenamefont {Pagano},
  \citenamefont {You}, \citenamefont {Gorshkov},\ and\ \citenamefont
  {Monroe}}]{Morong2021}%
  \BibitemOpen
  \bibfield  {author} {\bibinfo {author} {\bibfnamefont {W.}~\bibnamefont
  {Morong}}, \bibinfo {author} {\bibfnamefont {F.}~\bibnamefont {Liu}},
  \bibinfo {author} {\bibfnamefont {P.}~\bibnamefont {Becker}}, \bibinfo
  {author} {\bibfnamefont {K.~S.}\ \bibnamefont {Collins}}, \bibinfo {author}
  {\bibfnamefont {L.}~\bibnamefont {Feng}}, \bibinfo {author} {\bibfnamefont
  {A.}~\bibnamefont {Kyprianidis}}, \bibinfo {author} {\bibfnamefont
  {G.}~\bibnamefont {Pagano}}, \bibinfo {author} {\bibfnamefont
  {T.}~\bibnamefont {You}}, \bibinfo {author} {\bibfnamefont {A.~V.}\
  \bibnamefont {Gorshkov}},\ and\ \bibinfo {author} {\bibfnamefont
  {C.}~\bibnamefont {Monroe}},\ }\bibfield  {title} {\bibinfo {title}
  {{Observation of Stark many-body localization without disorder}},\ }\href
  {https://doi.org/10.1038/s41586-021-03988-0} {\bibfield  {journal} {\bibinfo
  {journal} {Nature}\ }\textbf {\bibinfo {volume} {599}},\ \bibinfo {pages}
  {393} (\bibinfo {year} {2021})}\BibitemShut {NoStop}%
\bibitem [{\citenamefont {Bloch}\ \emph {et~al.}(2012)\citenamefont {Bloch},
  \citenamefont {Dalibard},\ and\ \citenamefont
  {Nascimb{\`{e}}ne}}]{Bloch2012}%
  \BibitemOpen
  \bibfield  {author} {\bibinfo {author} {\bibfnamefont {I.}~\bibnamefont
  {Bloch}}, \bibinfo {author} {\bibfnamefont {J.}~\bibnamefont {Dalibard}},\
  and\ \bibinfo {author} {\bibfnamefont {S.}~\bibnamefont {Nascimb{\`{e}}ne}},\
  }\bibfield  {title} {\bibinfo {title} {{Quantum simulations with ultracold
  quantum gases}},\ }\href {https://doi.org/10.1038/nphys2259} {\bibfield
  {journal} {\bibinfo  {journal} {Nat. Phys.}\ }\textbf {\bibinfo {volume}
  {8}},\ \bibinfo {pages} {267} (\bibinfo {year} {2012})}\BibitemShut {NoStop}%
\bibitem [{\citenamefont {Gross}\ and\ \citenamefont
  {Bloch}(2017)}]{Gross2017}%
  \BibitemOpen
  \bibfield  {author} {\bibinfo {author} {\bibfnamefont {C.}~\bibnamefont
  {Gross}}\ and\ \bibinfo {author} {\bibfnamefont {I.}~\bibnamefont {Bloch}},\
  }\bibfield  {title} {\bibinfo {title} {{Quantum simulations with ultracold
  atoms in optical lattices}},\ }\href
  {https://doi.org/10.1126/science.aal3837} {\bibfield  {journal} {\bibinfo
  {journal} {Science}\ }\textbf {\bibinfo {volume} {357}},\ \bibinfo {pages}
  {995} (\bibinfo {year} {2017})}\BibitemShut {NoStop}%
\bibitem [{\citenamefont {Saffman}(2016)}]{Saffman2016}%
  \BibitemOpen
  \bibfield  {author} {\bibinfo {author} {\bibfnamefont {M.}~\bibnamefont
  {Saffman}},\ }\bibfield  {title} {\bibinfo {title} {{Quantum computing with
  atomic qubits and Rydberg interactions: progress and challenges}},\ }\href
  {https://doi.org/10.1088/0953-4075/49/20/202001} {\bibfield  {journal}
  {\bibinfo  {journal} {J. Phys. B At. Mol. Opt. Phys.}\ }\textbf {\bibinfo
  {volume} {49}},\ \bibinfo {pages} {202001} (\bibinfo {year}
  {2016})}\BibitemShut {NoStop}%
\bibitem [{\citenamefont {Bernien}\ \emph {et~al.}(2017)\citenamefont
  {Bernien}, \citenamefont {Schwartz}, \citenamefont {Keesling}, \citenamefont
  {Levine}, \citenamefont {Omran}, \citenamefont {Pichler}, \citenamefont
  {Choi}, \citenamefont {Zibrov}, \citenamefont {Endres}, \citenamefont
  {Greiner}, \citenamefont {Vuletic},\ and\ \citenamefont
  {Lukin}}]{Bernien2017}%
  \BibitemOpen
  \bibfield  {author} {\bibinfo {author} {\bibfnamefont {H.}~\bibnamefont
  {Bernien}}, \bibinfo {author} {\bibfnamefont {S.}~\bibnamefont {Schwartz}},
  \bibinfo {author} {\bibfnamefont {A.}~\bibnamefont {Keesling}}, \bibinfo
  {author} {\bibfnamefont {H.}~\bibnamefont {Levine}}, \bibinfo {author}
  {\bibfnamefont {A.}~\bibnamefont {Omran}}, \bibinfo {author} {\bibfnamefont
  {H.}~\bibnamefont {Pichler}}, \bibinfo {author} {\bibfnamefont
  {S.}~\bibnamefont {Choi}}, \bibinfo {author} {\bibfnamefont {A.~S.}\
  \bibnamefont {Zibrov}}, \bibinfo {author} {\bibfnamefont {M.}~\bibnamefont
  {Endres}}, \bibinfo {author} {\bibfnamefont {M.}~\bibnamefont {Greiner}},
  \bibinfo {author} {\bibfnamefont {V.}~\bibnamefont {Vuletic}},\ and\ \bibinfo
  {author} {\bibfnamefont {M.~D.}\ \bibnamefont {Lukin}},\ }\bibfield  {title}
  {\bibinfo {title} {{Probing many-body dynamics on a 51-atom quantum
  simulator}},\ }\href {https://doi.org/10.1038/nature24622} {\bibfield
  {journal} {\bibinfo  {journal} {Nature}\ }\textbf {\bibinfo {volume} {551}},\
  \bibinfo {pages} {579} (\bibinfo {year} {2017})}\BibitemShut {NoStop}%
\bibitem [{\citenamefont {Makhlin}\ \emph {et~al.}(2001)\citenamefont
  {Makhlin}, \citenamefont {Sch{\"{o}}n},\ and\ \citenamefont
  {Shnirman}}]{Makhlin2001}%
  \BibitemOpen
  \bibfield  {author} {\bibinfo {author} {\bibfnamefont {Y.}~\bibnamefont
  {Makhlin}}, \bibinfo {author} {\bibfnamefont {G.}~\bibnamefont
  {Sch{\"{o}}n}},\ and\ \bibinfo {author} {\bibfnamefont {A.}~\bibnamefont
  {Shnirman}},\ }\bibfield  {title} {\bibinfo {title} {{Quantum-state
  engineering with Josephson-junction devices}},\ }\href
  {https://doi.org/10.1103/RevModPhys.73.357} {\bibfield  {journal} {\bibinfo
  {journal} {Rev. Mod. Phys.}\ }\textbf {\bibinfo {volume} {73}},\ \bibinfo
  {pages} {357} (\bibinfo {year} {2001})}\BibitemShut {NoStop}%
\bibitem [{\citenamefont {Xiang}\ \emph {et~al.}(2013)\citenamefont {Xiang},
  \citenamefont {Ashhab}, \citenamefont {You},\ and\ \citenamefont
  {Nori}}]{Xiang2013}%
  \BibitemOpen
  \bibfield  {author} {\bibinfo {author} {\bibfnamefont {Z.-L.}\ \bibnamefont
  {Xiang}}, \bibinfo {author} {\bibfnamefont {S.}~\bibnamefont {Ashhab}},
  \bibinfo {author} {\bibfnamefont {J.~Q.}\ \bibnamefont {You}},\ and\ \bibinfo
  {author} {\bibfnamefont {F.}~\bibnamefont {Nori}},\ }\bibfield  {title}
  {\bibinfo {title} {{Hybrid quantum circuits: Superconducting circuits
  interacting with other quantum systems}},\ }\href
  {https://doi.org/10.1103/RevModPhys.85.623} {\bibfield  {journal} {\bibinfo
  {journal} {Rev. Mod. Phys.}\ }\textbf {\bibinfo {volume} {85}},\ \bibinfo
  {pages} {623} (\bibinfo {year} {2013})}\BibitemShut {NoStop}%
\bibitem [{\citenamefont {Gu}\ \emph {et~al.}(2017)\citenamefont {Gu},
  \citenamefont {Kockum}, \citenamefont {Miranowicz}, \citenamefont {Liu},\
  and\ \citenamefont {Nori}}]{Gu2017}%
  \BibitemOpen
  \bibfield  {author} {\bibinfo {author} {\bibfnamefont {X.}~\bibnamefont
  {Gu}}, \bibinfo {author} {\bibfnamefont {A.~F.}\ \bibnamefont {Kockum}},
  \bibinfo {author} {\bibfnamefont {A.}~\bibnamefont {Miranowicz}}, \bibinfo
  {author} {\bibfnamefont {Y.-x.}\ \bibnamefont {Liu}},\ and\ \bibinfo {author}
  {\bibfnamefont {F.}~\bibnamefont {Nori}},\ }\bibfield  {title} {\bibinfo
  {title} {{Microwave photonics with superconducting quantum circuits}},\
  }\href {https://doi.org/10.1016/j.physrep.2017.10.002} {\bibfield  {journal}
  {\bibinfo  {journal} {Phys. Rep.}\ }\textbf {\bibinfo {volume} {718-719}},\
  \bibinfo {pages} {1} (\bibinfo {year} {2017})}\BibitemShut {NoStop}%
\bibitem [{\citenamefont {Vandersypen}\ and\ \citenamefont
  {Chuang}(2005)}]{vandersypen2005nmr}%
  \BibitemOpen
  \bibfield  {author} {\bibinfo {author} {\bibfnamefont {L.~M.}\ \bibnamefont
  {Vandersypen}}\ and\ \bibinfo {author} {\bibfnamefont {I.~L.}\ \bibnamefont
  {Chuang}},\ }\bibfield  {title} {\bibinfo {title} {Nmr techniques for quantum
  control and computation},\ }\href@noop {} {\bibfield  {journal} {\bibinfo
  {journal} {Reviews of modern physics}\ }\textbf {\bibinfo {volume} {76}},\
  \bibinfo {pages} {1037} (\bibinfo {year} {2005})}\BibitemShut {NoStop}%
\bibitem [{\citenamefont {Rovny}\ \emph {et~al.}(2018)\citenamefont {Rovny},
  \citenamefont {Blum},\ and\ \citenamefont {Barrett}}]{Rovny2018}%
  \BibitemOpen
  \bibfield  {author} {\bibinfo {author} {\bibfnamefont {J.}~\bibnamefont
  {Rovny}}, \bibinfo {author} {\bibfnamefont {R.~L.}\ \bibnamefont {Blum}},\
  and\ \bibinfo {author} {\bibfnamefont {S.~E.}\ \bibnamefont {Barrett}},\
  }\bibfield  {title} {\bibinfo {title} {{Observation of Discrete-Time-Crystal
  Signatures in an Ordered Dipolar Many-Body System}},\ }\href
  {https://doi.org/10.1103/PhysRevLett.120.180603} {\bibfield  {journal}
  {\bibinfo  {journal} {Phys. Rev. Lett.}\ }\textbf {\bibinfo {volume} {120}},\
  \bibinfo {pages} {180603} (\bibinfo {year} {2018})}\BibitemShut {NoStop}%
\bibitem [{\citenamefont {Chen}\ \emph {et~al.}(2021)\citenamefont {Chen},
  \citenamefont {Sun}, \citenamefont {Gong}, \citenamefont {Zhu}, \citenamefont
  {Zhang}, \citenamefont {Wu}, \citenamefont {Ye}, \citenamefont {Zha},
  \citenamefont {Li}, \citenamefont {Guo}, \citenamefont {Qian}, \citenamefont
  {Huang}, \citenamefont {Yu}, \citenamefont {Deng}, \citenamefont {Rong},
  \citenamefont {Lin}, \citenamefont {Xu}, \citenamefont {Sun}, \citenamefont
  {Guo}, \citenamefont {Li}, \citenamefont {Liang}, \citenamefont {Peng},
  \citenamefont {Fan}, \citenamefont {Zhu},\ and\ \citenamefont
  {Pan}}]{Chen2021}%
  \BibitemOpen
  \bibfield  {author} {\bibinfo {author} {\bibfnamefont {F.}~\bibnamefont
  {Chen}}, \bibinfo {author} {\bibfnamefont {Z.-H.}\ \bibnamefont {Sun}},
  \bibinfo {author} {\bibfnamefont {M.}~\bibnamefont {Gong}}, \bibinfo {author}
  {\bibfnamefont {Q.}~\bibnamefont {Zhu}}, \bibinfo {author} {\bibfnamefont
  {Y.-R.}\ \bibnamefont {Zhang}}, \bibinfo {author} {\bibfnamefont
  {Y.}~\bibnamefont {Wu}}, \bibinfo {author} {\bibfnamefont {Y.}~\bibnamefont
  {Ye}}, \bibinfo {author} {\bibfnamefont {C.}~\bibnamefont {Zha}}, \bibinfo
  {author} {\bibfnamefont {S.}~\bibnamefont {Li}}, \bibinfo {author}
  {\bibfnamefont {S.}~\bibnamefont {Guo}}, \bibinfo {author} {\bibfnamefont
  {H.}~\bibnamefont {Qian}}, \bibinfo {author} {\bibfnamefont {H.-L.}\
  \bibnamefont {Huang}}, \bibinfo {author} {\bibfnamefont {J.}~\bibnamefont
  {Yu}}, \bibinfo {author} {\bibfnamefont {H.}~\bibnamefont {Deng}}, \bibinfo
  {author} {\bibfnamefont {H.}~\bibnamefont {Rong}}, \bibinfo {author}
  {\bibfnamefont {J.}~\bibnamefont {Lin}}, \bibinfo {author} {\bibfnamefont
  {Y.}~\bibnamefont {Xu}}, \bibinfo {author} {\bibfnamefont {L.}~\bibnamefont
  {Sun}}, \bibinfo {author} {\bibfnamefont {C.}~\bibnamefont {Guo}}, \bibinfo
  {author} {\bibfnamefont {N.}~\bibnamefont {Li}}, \bibinfo {author}
  {\bibfnamefont {F.}~\bibnamefont {Liang}}, \bibinfo {author} {\bibfnamefont
  {C.-Z.}\ \bibnamefont {Peng}}, \bibinfo {author} {\bibfnamefont
  {H.}~\bibnamefont {Fan}}, \bibinfo {author} {\bibfnamefont {X.}~\bibnamefont
  {Zhu}},\ and\ \bibinfo {author} {\bibfnamefont {J.-W.}\ \bibnamefont {Pan}},\
  }\bibfield  {title} {\bibinfo {title} {{Observation of Strong and Weak
  Thermalization in a Superconducting Quantum Processor}},\ }\href
  {https://doi.org/10.1103/PhysRevLett.127.020602} {\bibfield  {journal}
  {\bibinfo  {journal} {Phys. Rev. Lett.}\ }\textbf {\bibinfo {volume} {127}},\
  \bibinfo {pages} {020602} (\bibinfo {year} {2021})}\BibitemShut {NoStop}%
\bibitem [{\citenamefont {Roushan}\ \emph {et~al.}(2017)\citenamefont
  {Roushan}, \citenamefont {Neill}, \citenamefont {Tangpanitanon},
  \citenamefont {Bastidas}, \citenamefont {Megrant}, \citenamefont {Barends},
  \citenamefont {Chen}, \citenamefont {Chen}, \citenamefont {Chiaro},
  \citenamefont {Dunsworth}, \citenamefont {Fowler}, \citenamefont {Foxen},
  \citenamefont {Giustina}, \citenamefont {Jeffrey}, \citenamefont {Kelly},
  \citenamefont {Lucero}, \citenamefont {Mutus}, \citenamefont {Neeley},
  \citenamefont {Quintana}, \citenamefont {Sank}, \citenamefont {Vainsencher},
  \citenamefont {Wenner}, \citenamefont {White}, \citenamefont {Neven},
  \citenamefont {Angelakis},\ and\ \citenamefont {Martinis}}]{Roushan2017}%
  \BibitemOpen
  \bibfield  {author} {\bibinfo {author} {\bibfnamefont {P.}~\bibnamefont
  {Roushan}}, \bibinfo {author} {\bibfnamefont {C.}~\bibnamefont {Neill}},
  \bibinfo {author} {\bibfnamefont {J.}~\bibnamefont {Tangpanitanon}}, \bibinfo
  {author} {\bibfnamefont {V.~M.}\ \bibnamefont {Bastidas}}, \bibinfo {author}
  {\bibfnamefont {A.}~\bibnamefont {Megrant}}, \bibinfo {author} {\bibfnamefont
  {R.}~\bibnamefont {Barends}}, \bibinfo {author} {\bibfnamefont
  {Y.}~\bibnamefont {Chen}}, \bibinfo {author} {\bibfnamefont {Z.}~\bibnamefont
  {Chen}}, \bibinfo {author} {\bibfnamefont {B.}~\bibnamefont {Chiaro}},
  \bibinfo {author} {\bibfnamefont {A.}~\bibnamefont {Dunsworth}}, \bibinfo
  {author} {\bibfnamefont {A.}~\bibnamefont {Fowler}}, \bibinfo {author}
  {\bibfnamefont {B.}~\bibnamefont {Foxen}}, \bibinfo {author} {\bibfnamefont
  {M.}~\bibnamefont {Giustina}}, \bibinfo {author} {\bibfnamefont
  {E.}~\bibnamefont {Jeffrey}}, \bibinfo {author} {\bibfnamefont
  {J.}~\bibnamefont {Kelly}}, \bibinfo {author} {\bibfnamefont
  {E.}~\bibnamefont {Lucero}}, \bibinfo {author} {\bibfnamefont
  {J.}~\bibnamefont {Mutus}}, \bibinfo {author} {\bibfnamefont
  {M.}~\bibnamefont {Neeley}}, \bibinfo {author} {\bibfnamefont
  {C.}~\bibnamefont {Quintana}}, \bibinfo {author} {\bibfnamefont
  {D.}~\bibnamefont {Sank}}, \bibinfo {author} {\bibfnamefont {A.}~\bibnamefont
  {Vainsencher}}, \bibinfo {author} {\bibfnamefont {J.}~\bibnamefont {Wenner}},
  \bibinfo {author} {\bibfnamefont {T.}~\bibnamefont {White}}, \bibinfo
  {author} {\bibfnamefont {H.}~\bibnamefont {Neven}}, \bibinfo {author}
  {\bibfnamefont {D.~G.}\ \bibnamefont {Angelakis}},\ and\ \bibinfo {author}
  {\bibfnamefont {J.}~\bibnamefont {Martinis}},\ }\bibfield  {title} {\bibinfo
  {title} {{Spectroscopic signatures of localization with interacting photons
  in superconducting qubits}},\ }\href
  {https://doi.org/10.1126/science.aao1401} {\bibfield  {journal} {\bibinfo
  {journal} {Science}\ }\textbf {\bibinfo {volume} {358}},\ \bibinfo {pages}
  {1175} (\bibinfo {year} {2017})}\BibitemShut {NoStop}%
\bibitem [{\citenamefont {Xu}\ \emph {et~al.}(2018)\citenamefont {Xu},
  \citenamefont {Chen}, \citenamefont {Zeng}, \citenamefont {Zhang},
  \citenamefont {Song}, \citenamefont {Liu}, \citenamefont {Guo}, \citenamefont
  {Zhang}, \citenamefont {Xu}, \citenamefont {Deng}, \citenamefont {Huang},
  \citenamefont {Wang}, \citenamefont {Zhu}, \citenamefont {Zheng},\ and\
  \citenamefont {Fan}}]{Xu2018}%
  \BibitemOpen
  \bibfield  {author} {\bibinfo {author} {\bibfnamefont {K.}~\bibnamefont
  {Xu}}, \bibinfo {author} {\bibfnamefont {J.-J.}\ \bibnamefont {Chen}},
  \bibinfo {author} {\bibfnamefont {Y.}~\bibnamefont {Zeng}}, \bibinfo {author}
  {\bibfnamefont {Y.-R.}\ \bibnamefont {Zhang}}, \bibinfo {author}
  {\bibfnamefont {C.}~\bibnamefont {Song}}, \bibinfo {author} {\bibfnamefont
  {W.}~\bibnamefont {Liu}}, \bibinfo {author} {\bibfnamefont {Q.}~\bibnamefont
  {Guo}}, \bibinfo {author} {\bibfnamefont {P.}~\bibnamefont {Zhang}}, \bibinfo
  {author} {\bibfnamefont {D.}~\bibnamefont {Xu}}, \bibinfo {author}
  {\bibfnamefont {H.}~\bibnamefont {Deng}}, \bibinfo {author} {\bibfnamefont
  {K.}~\bibnamefont {Huang}}, \bibinfo {author} {\bibfnamefont
  {H.}~\bibnamefont {Wang}}, \bibinfo {author} {\bibfnamefont {X.}~\bibnamefont
  {Zhu}}, \bibinfo {author} {\bibfnamefont {D.}~\bibnamefont {Zheng}},\ and\
  \bibinfo {author} {\bibfnamefont {H.}~\bibnamefont {Fan}},\ }\bibfield
  {title} {\bibinfo {title} {{Emulating Many-Body Localization with a
  Superconducting Quantum Processor}},\ }\href
  {https://doi.org/10.1103/PhysRevLett.120.050507} {\bibfield  {journal}
  {\bibinfo  {journal} {Phys. Rev. Lett.}\ }\textbf {\bibinfo {volume} {120}},\
  \bibinfo {pages} {050507} (\bibinfo {year} {2018})}\BibitemShut {NoStop}%
\bibitem [{\citenamefont {Guo}\ \emph {et~al.}(2021{\natexlab{a}})\citenamefont
  {Guo}, \citenamefont {Cheng}, \citenamefont {Li}, \citenamefont {Xu},
  \citenamefont {Zhang}, \citenamefont {Wang}, \citenamefont {Song},
  \citenamefont {Liu}, \citenamefont {Ren}, \citenamefont {Dong}, \citenamefont
  {Mondaini},\ and\ \citenamefont {Wang}}]{Guo2020}%
  \BibitemOpen
  \bibfield  {author} {\bibinfo {author} {\bibfnamefont {Q.}~\bibnamefont
  {Guo}}, \bibinfo {author} {\bibfnamefont {C.}~\bibnamefont {Cheng}}, \bibinfo
  {author} {\bibfnamefont {H.}~\bibnamefont {Li}}, \bibinfo {author}
  {\bibfnamefont {S.}~\bibnamefont {Xu}}, \bibinfo {author} {\bibfnamefont
  {P.}~\bibnamefont {Zhang}}, \bibinfo {author} {\bibfnamefont
  {Z.}~\bibnamefont {Wang}}, \bibinfo {author} {\bibfnamefont {C.}~\bibnamefont
  {Song}}, \bibinfo {author} {\bibfnamefont {W.}~\bibnamefont {Liu}}, \bibinfo
  {author} {\bibfnamefont {W.}~\bibnamefont {Ren}}, \bibinfo {author}
  {\bibfnamefont {H.}~\bibnamefont {Dong}}, \bibinfo {author} {\bibfnamefont
  {R.}~\bibnamefont {Mondaini}},\ and\ \bibinfo {author} {\bibfnamefont
  {H.}~\bibnamefont {Wang}},\ }\bibfield  {title} {\bibinfo {title} {{Stark
  Many-Body Localization on a Superconducting Quantum Processor}},\ }\href
  {https://doi.org/10.1103/PhysRevLett.127.240502} {\bibfield  {journal}
  {\bibinfo  {journal} {Phys. Rev. Lett.}\ }\textbf {\bibinfo {volume} {127}},\
  \bibinfo {pages} {240502} (\bibinfo {year} {2021}{\natexlab{a}})}\BibitemShut
  {NoStop}%
\bibitem [{\citenamefont {Guo}\ \emph {et~al.}(2021{\natexlab{b}})\citenamefont
  {Guo}, \citenamefont {Cheng}, \citenamefont {Sun}, \citenamefont {Song},
  \citenamefont {Li}, \citenamefont {Wang}, \citenamefont {Ren}, \citenamefont
  {Dong}, \citenamefont {Zheng}, \citenamefont {Zhang}, \citenamefont
  {Mondaini}, \citenamefont {Fan},\ and\ \citenamefont {Wang}}]{Guo2021}%
  \BibitemOpen
  \bibfield  {author} {\bibinfo {author} {\bibfnamefont {Q.}~\bibnamefont
  {Guo}}, \bibinfo {author} {\bibfnamefont {C.}~\bibnamefont {Cheng}}, \bibinfo
  {author} {\bibfnamefont {Z.-H.}\ \bibnamefont {Sun}}, \bibinfo {author}
  {\bibfnamefont {Z.}~\bibnamefont {Song}}, \bibinfo {author} {\bibfnamefont
  {H.}~\bibnamefont {Li}}, \bibinfo {author} {\bibfnamefont {Z.}~\bibnamefont
  {Wang}}, \bibinfo {author} {\bibfnamefont {W.}~\bibnamefont {Ren}}, \bibinfo
  {author} {\bibfnamefont {H.}~\bibnamefont {Dong}}, \bibinfo {author}
  {\bibfnamefont {D.}~\bibnamefont {Zheng}}, \bibinfo {author} {\bibfnamefont
  {Y.-r.}\ \bibnamefont {Zhang}}, \bibinfo {author} {\bibfnamefont
  {R.}~\bibnamefont {Mondaini}}, \bibinfo {author} {\bibfnamefont
  {H.}~\bibnamefont {Fan}},\ and\ \bibinfo {author} {\bibfnamefont
  {H.}~\bibnamefont {Wang}},\ }\bibfield  {title} {\bibinfo {title}
  {{Observation of energy-resolved many-body localization}},\ }\href
  {https://doi.org/10.1038/s41567-020-1035-1} {\bibfield  {journal} {\bibinfo
  {journal} {Nat. Phys.}\ }\textbf {\bibinfo {volume} {17}},\ \bibinfo {pages}
  {234} (\bibinfo {year} {2021}{\natexlab{b}})}\BibitemShut {NoStop}%
\bibitem [{\citenamefont {Mi}\ \emph {et~al.}(2022)\citenamefont {Mi},
  \citenamefont {Ippoliti}, \citenamefont {Quintana}, \citenamefont {Greene},
  \citenamefont {Chen}, \citenamefont {Gross}, \citenamefont {Arute},
  \citenamefont {Arya}, \citenamefont {Atalaya}, \citenamefont {Babbush},
  \citenamefont {Bardin}, \citenamefont {Basso}, \citenamefont {Bengtsson},
  \citenamefont {Bilmes}, \citenamefont {Bourassa}, \citenamefont {Brill},
  \citenamefont {Broughton}, \citenamefont {Buckley}, \citenamefont {Buell},
  \citenamefont {Burkett}, \citenamefont {Bushnell}, \citenamefont {Chiaro},
  \citenamefont {Collins}, \citenamefont {Courtney}, \citenamefont {Debroy},
  \citenamefont {Demura}, \citenamefont {Derk}, \citenamefont {Dunsworth},
  \citenamefont {Eppens}, \citenamefont {Erickson}, \citenamefont {Farhi},
  \citenamefont {Fowler}, \citenamefont {Foxen}, \citenamefont {Gidney},
  \citenamefont {Giustina}, \citenamefont {Harrigan}, \citenamefont
  {Harrington}, \citenamefont {Hilton}, \citenamefont {Ho}, \citenamefont
  {Hong}, \citenamefont {Huang}, \citenamefont {Huff}, \citenamefont {Huggins},
  \citenamefont {Ioffe}, \citenamefont {Isakov}, \citenamefont {Iveland},
  \citenamefont {Jeffrey}, \citenamefont {Jiang}, \citenamefont {Jones},
  \citenamefont {Kafri}, \citenamefont {Khattar}, \citenamefont {Kim},
  \citenamefont {Kitaev}, \citenamefont {Klimov}, \citenamefont {Korotkov},
  \citenamefont {Kostritsa}, \citenamefont {Landhuis}, \citenamefont {Laptev},
  \citenamefont {Lee}, \citenamefont {Lee}, \citenamefont {Locharla},
  \citenamefont {Lucero}, \citenamefont {Martin}, \citenamefont {McClean},
  \citenamefont {McCourt}, \citenamefont {McEwen}, \citenamefont {Miao},
  \citenamefont {Mohseni}, \citenamefont {Montazeri}, \citenamefont
  {Mruczkiewicz}, \citenamefont {Naaman}, \citenamefont {Neeley}, \citenamefont
  {Neill}, \citenamefont {Newman}, \citenamefont {Niu}, \citenamefont
  {O'Brien}, \citenamefont {Opremcak}, \citenamefont {Ostby}, \citenamefont
  {Pato}, \citenamefont {Petukhov}, \citenamefont {Rubin}, \citenamefont
  {Sank}, \citenamefont {Satzinger}, \citenamefont {Shvarts}, \citenamefont
  {Su}, \citenamefont {Strain}, \citenamefont {Szalay}, \citenamefont
  {Trevithick}, \citenamefont {Villalonga}, \citenamefont {White},
  \citenamefont {Yao}, \citenamefont {Yeh}, \citenamefont {Yoo}, \citenamefont
  {Zalcman}, \citenamefont {Neven}, \citenamefont {Boixo}, \citenamefont
  {Smelyanskiy}, \citenamefont {Megrant}, \citenamefont {Kelly}, \citenamefont
  {Chen}, \citenamefont {Sondhi}, \citenamefont {Moessner}, \citenamefont
  {Kechedzhi}, \citenamefont {Khemani},\ and\ \citenamefont
  {Roushan}}]{Mi2022}%
  \BibitemOpen
  \bibfield  {author} {\bibinfo {author} {\bibfnamefont {X.}~\bibnamefont
  {Mi}}, \bibinfo {author} {\bibfnamefont {M.}~\bibnamefont {Ippoliti}},
  \bibinfo {author} {\bibfnamefont {C.}~\bibnamefont {Quintana}}, \bibinfo
  {author} {\bibfnamefont {A.}~\bibnamefont {Greene}}, \bibinfo {author}
  {\bibfnamefont {Z.}~\bibnamefont {Chen}}, \bibinfo {author} {\bibfnamefont
  {J.}~\bibnamefont {Gross}}, \bibinfo {author} {\bibfnamefont
  {F.}~\bibnamefont {Arute}}, \bibinfo {author} {\bibfnamefont
  {K.}~\bibnamefont {Arya}}, \bibinfo {author} {\bibfnamefont {J.}~\bibnamefont
  {Atalaya}}, \bibinfo {author} {\bibfnamefont {R.}~\bibnamefont {Babbush}},
  \bibinfo {author} {\bibfnamefont {J.~C.}\ \bibnamefont {Bardin}}, \bibinfo
  {author} {\bibfnamefont {J.}~\bibnamefont {Basso}}, \bibinfo {author}
  {\bibfnamefont {A.}~\bibnamefont {Bengtsson}}, \bibinfo {author}
  {\bibfnamefont {A.}~\bibnamefont {Bilmes}}, \bibinfo {author} {\bibfnamefont
  {A.}~\bibnamefont {Bourassa}}, \bibinfo {author} {\bibfnamefont
  {L.}~\bibnamefont {Brill}}, \bibinfo {author} {\bibfnamefont
  {M.}~\bibnamefont {Broughton}}, \bibinfo {author} {\bibfnamefont {B.~B.}\
  \bibnamefont {Buckley}}, \bibinfo {author} {\bibfnamefont {D.~A.}\
  \bibnamefont {Buell}}, \bibinfo {author} {\bibfnamefont {B.}~\bibnamefont
  {Burkett}}, \bibinfo {author} {\bibfnamefont {N.}~\bibnamefont {Bushnell}},
  \bibinfo {author} {\bibfnamefont {B.}~\bibnamefont {Chiaro}}, \bibinfo
  {author} {\bibfnamefont {R.}~\bibnamefont {Collins}}, \bibinfo {author}
  {\bibfnamefont {W.}~\bibnamefont {Courtney}}, \bibinfo {author}
  {\bibfnamefont {D.}~\bibnamefont {Debroy}}, \bibinfo {author} {\bibfnamefont
  {S.}~\bibnamefont {Demura}}, \bibinfo {author} {\bibfnamefont {A.~R.}\
  \bibnamefont {Derk}}, \bibinfo {author} {\bibfnamefont {A.}~\bibnamefont
  {Dunsworth}}, \bibinfo {author} {\bibfnamefont {D.}~\bibnamefont {Eppens}},
  \bibinfo {author} {\bibfnamefont {C.}~\bibnamefont {Erickson}}, \bibinfo
  {author} {\bibfnamefont {E.}~\bibnamefont {Farhi}}, \bibinfo {author}
  {\bibfnamefont {A.~G.}\ \bibnamefont {Fowler}}, \bibinfo {author}
  {\bibfnamefont {B.}~\bibnamefont {Foxen}}, \bibinfo {author} {\bibfnamefont
  {C.}~\bibnamefont {Gidney}}, \bibinfo {author} {\bibfnamefont
  {M.}~\bibnamefont {Giustina}}, \bibinfo {author} {\bibfnamefont {M.~P.}\
  \bibnamefont {Harrigan}}, \bibinfo {author} {\bibfnamefont {S.~D.}\
  \bibnamefont {Harrington}}, \bibinfo {author} {\bibfnamefont
  {J.}~\bibnamefont {Hilton}}, \bibinfo {author} {\bibfnamefont
  {A.}~\bibnamefont {Ho}}, \bibinfo {author} {\bibfnamefont {S.}~\bibnamefont
  {Hong}}, \bibinfo {author} {\bibfnamefont {T.}~\bibnamefont {Huang}},
  \bibinfo {author} {\bibfnamefont {A.}~\bibnamefont {Huff}}, \bibinfo {author}
  {\bibfnamefont {W.~J.}\ \bibnamefont {Huggins}}, \bibinfo {author}
  {\bibfnamefont {L.~B.}\ \bibnamefont {Ioffe}}, \bibinfo {author}
  {\bibfnamefont {S.~V.}\ \bibnamefont {Isakov}}, \bibinfo {author}
  {\bibfnamefont {J.}~\bibnamefont {Iveland}}, \bibinfo {author} {\bibfnamefont
  {E.}~\bibnamefont {Jeffrey}}, \bibinfo {author} {\bibfnamefont
  {Z.}~\bibnamefont {Jiang}}, \bibinfo {author} {\bibfnamefont
  {C.}~\bibnamefont {Jones}}, \bibinfo {author} {\bibfnamefont
  {D.}~\bibnamefont {Kafri}}, \bibinfo {author} {\bibfnamefont
  {T.}~\bibnamefont {Khattar}}, \bibinfo {author} {\bibfnamefont
  {S.}~\bibnamefont {Kim}}, \bibinfo {author} {\bibfnamefont {A.}~\bibnamefont
  {Kitaev}}, \bibinfo {author} {\bibfnamefont {P.~V.}\ \bibnamefont {Klimov}},
  \bibinfo {author} {\bibfnamefont {A.~N.}\ \bibnamefont {Korotkov}}, \bibinfo
  {author} {\bibfnamefont {F.}~\bibnamefont {Kostritsa}}, \bibinfo {author}
  {\bibfnamefont {D.}~\bibnamefont {Landhuis}}, \bibinfo {author}
  {\bibfnamefont {P.}~\bibnamefont {Laptev}}, \bibinfo {author} {\bibfnamefont
  {J.}~\bibnamefont {Lee}}, \bibinfo {author} {\bibfnamefont {K.}~\bibnamefont
  {Lee}}, \bibinfo {author} {\bibfnamefont {A.}~\bibnamefont {Locharla}},
  \bibinfo {author} {\bibfnamefont {E.}~\bibnamefont {Lucero}}, \bibinfo
  {author} {\bibfnamefont {O.}~\bibnamefont {Martin}}, \bibinfo {author}
  {\bibfnamefont {J.~R.}\ \bibnamefont {McClean}}, \bibinfo {author}
  {\bibfnamefont {T.}~\bibnamefont {McCourt}}, \bibinfo {author} {\bibfnamefont
  {M.}~\bibnamefont {McEwen}}, \bibinfo {author} {\bibfnamefont {K.~C.}\
  \bibnamefont {Miao}}, \bibinfo {author} {\bibfnamefont {M.}~\bibnamefont
  {Mohseni}}, \bibinfo {author} {\bibfnamefont {S.}~\bibnamefont {Montazeri}},
  \bibinfo {author} {\bibfnamefont {W.}~\bibnamefont {Mruczkiewicz}}, \bibinfo
  {author} {\bibfnamefont {O.}~\bibnamefont {Naaman}}, \bibinfo {author}
  {\bibfnamefont {M.}~\bibnamefont {Neeley}}, \bibinfo {author} {\bibfnamefont
  {C.}~\bibnamefont {Neill}}, \bibinfo {author} {\bibfnamefont
  {M.}~\bibnamefont {Newman}}, \bibinfo {author} {\bibfnamefont {M.~Y.}\
  \bibnamefont {Niu}}, \bibinfo {author} {\bibfnamefont {T.~E.}\ \bibnamefont
  {O'Brien}}, \bibinfo {author} {\bibfnamefont {A.}~\bibnamefont {Opremcak}},
  \bibinfo {author} {\bibfnamefont {E.}~\bibnamefont {Ostby}}, \bibinfo
  {author} {\bibfnamefont {B.}~\bibnamefont {Pato}}, \bibinfo {author}
  {\bibfnamefont {A.}~\bibnamefont {Petukhov}}, \bibinfo {author}
  {\bibfnamefont {N.~C.}\ \bibnamefont {Rubin}}, \bibinfo {author}
  {\bibfnamefont {D.}~\bibnamefont {Sank}}, \bibinfo {author} {\bibfnamefont
  {K.~J.}\ \bibnamefont {Satzinger}}, \bibinfo {author} {\bibfnamefont
  {V.}~\bibnamefont {Shvarts}}, \bibinfo {author} {\bibfnamefont
  {Y.}~\bibnamefont {Su}}, \bibinfo {author} {\bibfnamefont {D.}~\bibnamefont
  {Strain}}, \bibinfo {author} {\bibfnamefont {M.}~\bibnamefont {Szalay}},
  \bibinfo {author} {\bibfnamefont {M.~D.}\ \bibnamefont {Trevithick}},
  \bibinfo {author} {\bibfnamefont {B.}~\bibnamefont {Villalonga}}, \bibinfo
  {author} {\bibfnamefont {T.}~\bibnamefont {White}}, \bibinfo {author}
  {\bibfnamefont {Z.~J.}\ \bibnamefont {Yao}}, \bibinfo {author} {\bibfnamefont
  {P.}~\bibnamefont {Yeh}}, \bibinfo {author} {\bibfnamefont {J.}~\bibnamefont
  {Yoo}}, \bibinfo {author} {\bibfnamefont {A.}~\bibnamefont {Zalcman}},
  \bibinfo {author} {\bibfnamefont {H.}~\bibnamefont {Neven}}, \bibinfo
  {author} {\bibfnamefont {S.}~\bibnamefont {Boixo}}, \bibinfo {author}
  {\bibfnamefont {V.}~\bibnamefont {Smelyanskiy}}, \bibinfo {author}
  {\bibfnamefont {A.}~\bibnamefont {Megrant}}, \bibinfo {author} {\bibfnamefont
  {J.}~\bibnamefont {Kelly}}, \bibinfo {author} {\bibfnamefont
  {Y.}~\bibnamefont {Chen}}, \bibinfo {author} {\bibfnamefont {S.~L.}\
  \bibnamefont {Sondhi}}, \bibinfo {author} {\bibfnamefont {R.}~\bibnamefont
  {Moessner}}, \bibinfo {author} {\bibfnamefont {K.}~\bibnamefont {Kechedzhi}},
  \bibinfo {author} {\bibfnamefont {V.}~\bibnamefont {Khemani}},\ and\ \bibinfo
  {author} {\bibfnamefont {P.}~\bibnamefont {Roushan}},\ }\bibfield  {title}
  {\bibinfo {title} {{Time-crystalline eigenstate order on a quantum
  processor}},\ }\href {https://doi.org/10.1038/s41586-021-04257-w} {\bibfield
  {journal} {\bibinfo  {journal} {Nature}\ }\textbf {\bibinfo {volume} {601}},\
  \bibinfo {pages} {531} (\bibinfo {year} {2022})}\BibitemShut {NoStop}%
\bibitem [{\citenamefont {Yan}\ \emph {et~al.}(2019)\citenamefont {Yan},
  \citenamefont {Zhang}, \citenamefont {Gong}, \citenamefont {Wu},
  \citenamefont {Zheng}, \citenamefont {Li}, \citenamefont {Wang},
  \citenamefont {Liang}, \citenamefont {Lin}, \citenamefont {Xu}, \citenamefont
  {Guo}, \citenamefont {Sun}, \citenamefont {Peng}, \citenamefont {Xia},
  \citenamefont {Deng}, \citenamefont {Rong}, \citenamefont {You},
  \citenamefont {Nori}, \citenamefont {Fan}, \citenamefont {Zhu},\ and\
  \citenamefont {Pan}}]{Yan2019}%
  \BibitemOpen
  \bibfield  {author} {\bibinfo {author} {\bibfnamefont {Z.}~\bibnamefont
  {Yan}}, \bibinfo {author} {\bibfnamefont {Y.-R.}\ \bibnamefont {Zhang}},
  \bibinfo {author} {\bibfnamefont {M.}~\bibnamefont {Gong}}, \bibinfo {author}
  {\bibfnamefont {Y.}~\bibnamefont {Wu}}, \bibinfo {author} {\bibfnamefont
  {Y.}~\bibnamefont {Zheng}}, \bibinfo {author} {\bibfnamefont
  {S.}~\bibnamefont {Li}}, \bibinfo {author} {\bibfnamefont {C.}~\bibnamefont
  {Wang}}, \bibinfo {author} {\bibfnamefont {F.}~\bibnamefont {Liang}},
  \bibinfo {author} {\bibfnamefont {J.}~\bibnamefont {Lin}}, \bibinfo {author}
  {\bibfnamefont {Y.}~\bibnamefont {Xu}}, \bibinfo {author} {\bibfnamefont
  {C.}~\bibnamefont {Guo}}, \bibinfo {author} {\bibfnamefont {L.}~\bibnamefont
  {Sun}}, \bibinfo {author} {\bibfnamefont {C.-Z.}\ \bibnamefont {Peng}},
  \bibinfo {author} {\bibfnamefont {K.}~\bibnamefont {Xia}}, \bibinfo {author}
  {\bibfnamefont {H.}~\bibnamefont {Deng}}, \bibinfo {author} {\bibfnamefont
  {H.}~\bibnamefont {Rong}}, \bibinfo {author} {\bibfnamefont {J.~Q.}\
  \bibnamefont {You}}, \bibinfo {author} {\bibfnamefont {F.}~\bibnamefont
  {Nori}}, \bibinfo {author} {\bibfnamefont {H.}~\bibnamefont {Fan}}, \bibinfo
  {author} {\bibfnamefont {X.}~\bibnamefont {Zhu}},\ and\ \bibinfo {author}
  {\bibfnamefont {J.-W.}\ \bibnamefont {Pan}},\ }\bibfield  {title} {\bibinfo
  {title} {{Strongly correlated quantum walks with a 12-qubit superconducting
  processor}},\ }\href {https://doi.org/10.1126/science.aaw1611} {\bibfield
  {journal} {\bibinfo  {journal} {Science}\ }\textbf {\bibinfo {volume}
  {364}},\ \bibinfo {pages} {753} (\bibinfo {year} {2019})}\BibitemShut
  {NoStop}%
\bibitem [{\citenamefont {Gong}\ \emph {et~al.}(2021)\citenamefont {Gong},
  \citenamefont {Wang}, \citenamefont {Zha}, \citenamefont {Chen},
  \citenamefont {Huang}, \citenamefont {Wu}, \citenamefont {Zhu}, \citenamefont
  {Zhao}, \citenamefont {Li}, \citenamefont {Guo}, \citenamefont {Qian},
  \citenamefont {Ye}, \citenamefont {Chen}, \citenamefont {Ying}, \citenamefont
  {Yu}, \citenamefont {Fan}, \citenamefont {Wu}, \citenamefont {Su},
  \citenamefont {Deng}, \citenamefont {Rong}, \citenamefont {Zhang},
  \citenamefont {Cao}, \citenamefont {Lin}, \citenamefont {Xu}, \citenamefont
  {Sun}, \citenamefont {Guo}, \citenamefont {Li}, \citenamefont {Liang},
  \citenamefont {Bastidas}, \citenamefont {Nemoto}, \citenamefont {Munro},
  \citenamefont {Huo}, \citenamefont {Lu}, \citenamefont {Peng}, \citenamefont
  {Zhu},\ and\ \citenamefont {Pan}}]{Gong2021}%
  \BibitemOpen
  \bibfield  {author} {\bibinfo {author} {\bibfnamefont {M.}~\bibnamefont
  {Gong}}, \bibinfo {author} {\bibfnamefont {S.}~\bibnamefont {Wang}}, \bibinfo
  {author} {\bibfnamefont {C.}~\bibnamefont {Zha}}, \bibinfo {author}
  {\bibfnamefont {M.-C.}\ \bibnamefont {Chen}}, \bibinfo {author}
  {\bibfnamefont {H.-L.}\ \bibnamefont {Huang}}, \bibinfo {author}
  {\bibfnamefont {Y.}~\bibnamefont {Wu}}, \bibinfo {author} {\bibfnamefont
  {Q.}~\bibnamefont {Zhu}}, \bibinfo {author} {\bibfnamefont {Y.}~\bibnamefont
  {Zhao}}, \bibinfo {author} {\bibfnamefont {S.}~\bibnamefont {Li}}, \bibinfo
  {author} {\bibfnamefont {S.}~\bibnamefont {Guo}}, \bibinfo {author}
  {\bibfnamefont {H.}~\bibnamefont {Qian}}, \bibinfo {author} {\bibfnamefont
  {Y.}~\bibnamefont {Ye}}, \bibinfo {author} {\bibfnamefont {F.}~\bibnamefont
  {Chen}}, \bibinfo {author} {\bibfnamefont {C.}~\bibnamefont {Ying}}, \bibinfo
  {author} {\bibfnamefont {J.}~\bibnamefont {Yu}}, \bibinfo {author}
  {\bibfnamefont {D.}~\bibnamefont {Fan}}, \bibinfo {author} {\bibfnamefont
  {D.}~\bibnamefont {Wu}}, \bibinfo {author} {\bibfnamefont {H.}~\bibnamefont
  {Su}}, \bibinfo {author} {\bibfnamefont {H.}~\bibnamefont {Deng}}, \bibinfo
  {author} {\bibfnamefont {H.}~\bibnamefont {Rong}}, \bibinfo {author}
  {\bibfnamefont {K.}~\bibnamefont {Zhang}}, \bibinfo {author} {\bibfnamefont
  {S.}~\bibnamefont {Cao}}, \bibinfo {author} {\bibfnamefont {J.}~\bibnamefont
  {Lin}}, \bibinfo {author} {\bibfnamefont {Y.}~\bibnamefont {Xu}}, \bibinfo
  {author} {\bibfnamefont {L.}~\bibnamefont {Sun}}, \bibinfo {author}
  {\bibfnamefont {C.}~\bibnamefont {Guo}}, \bibinfo {author} {\bibfnamefont
  {N.}~\bibnamefont {Li}}, \bibinfo {author} {\bibfnamefont {F.}~\bibnamefont
  {Liang}}, \bibinfo {author} {\bibfnamefont {V.~M.}\ \bibnamefont {Bastidas}},
  \bibinfo {author} {\bibfnamefont {K.}~\bibnamefont {Nemoto}}, \bibinfo
  {author} {\bibfnamefont {W.~J.}\ \bibnamefont {Munro}}, \bibinfo {author}
  {\bibfnamefont {Y.-H.}\ \bibnamefont {Huo}}, \bibinfo {author} {\bibfnamefont
  {C.-Y.}\ \bibnamefont {Lu}}, \bibinfo {author} {\bibfnamefont {C.-Z.}\
  \bibnamefont {Peng}}, \bibinfo {author} {\bibfnamefont {X.}~\bibnamefont
  {Zhu}},\ and\ \bibinfo {author} {\bibfnamefont {J.-W.}\ \bibnamefont {Pan}},\
  }\bibfield  {title} {\bibinfo {title} {{Quantum walks on a programmable
  two-dimensional 62-qubit superconducting processor}},\ }\href
  {https://doi.org/10.1126/science.abg7812} {\bibfield  {journal} {\bibinfo
  {journal} {Science}\ }\textbf {\bibinfo {volume} {372}},\ \bibinfo {pages}
  {948} (\bibinfo {year} {2021})}\BibitemShut {NoStop}%
\bibitem [{\citenamefont {Braum{\"{u}}ller}\ \emph {et~al.}(2022)\citenamefont
  {Braum{\"{u}}ller}, \citenamefont {Karamlou}, \citenamefont {Yanay},
  \citenamefont {Kannan}, \citenamefont {Kim}, \citenamefont {Kjaergaard},
  \citenamefont {Melville}, \citenamefont {Niedzielski}, \citenamefont {Sung},
  \citenamefont {Veps{\"{a}}l{\"{a}}inen}, \citenamefont {Winik}, \citenamefont
  {Yoder}, \citenamefont {Orlando}, \citenamefont {Gustavsson}, \citenamefont
  {Tahan},\ and\ \citenamefont {Oliver}}]{Braumuller2022}%
  \BibitemOpen
  \bibfield  {author} {\bibinfo {author} {\bibfnamefont {J.}~\bibnamefont
  {Braum{\"{u}}ller}}, \bibinfo {author} {\bibfnamefont {A.~H.}\ \bibnamefont
  {Karamlou}}, \bibinfo {author} {\bibfnamefont {Y.}~\bibnamefont {Yanay}},
  \bibinfo {author} {\bibfnamefont {B.}~\bibnamefont {Kannan}}, \bibinfo
  {author} {\bibfnamefont {D.}~\bibnamefont {Kim}}, \bibinfo {author}
  {\bibfnamefont {M.}~\bibnamefont {Kjaergaard}}, \bibinfo {author}
  {\bibfnamefont {A.}~\bibnamefont {Melville}}, \bibinfo {author}
  {\bibfnamefont {B.~M.}\ \bibnamefont {Niedzielski}}, \bibinfo {author}
  {\bibfnamefont {Y.}~\bibnamefont {Sung}}, \bibinfo {author} {\bibfnamefont
  {A.}~\bibnamefont {Veps{\"{a}}l{\"{a}}inen}}, \bibinfo {author}
  {\bibfnamefont {R.}~\bibnamefont {Winik}}, \bibinfo {author} {\bibfnamefont
  {J.~L.}\ \bibnamefont {Yoder}}, \bibinfo {author} {\bibfnamefont {T.~P.}\
  \bibnamefont {Orlando}}, \bibinfo {author} {\bibfnamefont {S.}~\bibnamefont
  {Gustavsson}}, \bibinfo {author} {\bibfnamefont {C.}~\bibnamefont {Tahan}},\
  and\ \bibinfo {author} {\bibfnamefont {W.~D.}\ \bibnamefont {Oliver}},\
  }\bibfield  {title} {\bibinfo {title} {{Probing quantum information
  propagation with out-of-time-ordered correlators}},\ }\href
  {https://doi.org/10.1038/s41567-021-01430-w} {\bibfield  {journal} {\bibinfo
  {journal} {Nat. Phys.}\ }\textbf {\bibinfo {volume} {18}},\ \bibinfo {pages}
  {172} (\bibinfo {year} {2022})}\BibitemShut {NoStop}%
\bibitem [{\citenamefont {Harper}(1955)}]{Harper1955}%
  \BibitemOpen
  \bibfield  {author} {\bibinfo {author} {\bibfnamefont {P.~G.}\ \bibnamefont
  {Harper}},\ }\bibfield  {title} {\bibinfo {title} {{Single band motion of
  conduction electrons in a uniform magnetic field}},\ }\href
  {https://doi.org/10.1088/0370-1298/68/10/304} {\bibfield  {journal} {\bibinfo
   {journal} {Proc. Phys. Soc. Sect. A}\ }\textbf {\bibinfo {volume} {68}},\
  \bibinfo {pages} {874} (\bibinfo {year} {1955})}\BibitemShut {NoStop}%
\bibitem [{\citenamefont {Aubry}\ and\ \citenamefont
  {Andr{\'e}}(1980)}]{aubry1980}%
  \BibitemOpen
  \bibfield  {author} {\bibinfo {author} {\bibfnamefont {S.}~\bibnamefont
  {Aubry}}\ and\ \bibinfo {author} {\bibfnamefont {G.}~\bibnamefont
  {Andr{\'e}}},\ }\bibfield  {title} {\bibinfo {title} {Analyticity breaking
  and anderson localization in incommensurate lattices},\ }\href@noop {}
  {\bibfield  {journal} {\bibinfo  {journal} {Ann. Israel Phys. Soc}\ }\textbf
  {\bibinfo {volume} {3}},\ \bibinfo {pages} {18} (\bibinfo {year}
  {1980})}\BibitemShut {NoStop}%
\bibitem [{\citenamefont {Thouless}(1983)}]{Thouless1983}%
  \BibitemOpen
  \bibfield  {author} {\bibinfo {author} {\bibfnamefont {D.~J.}\ \bibnamefont
  {Thouless}},\ }\bibfield  {title} {\bibinfo {title} {{Bandwidths for a
  quasiperiodic tight-binding model}},\ }\href
  {https://doi.org/10.1103/PhysRevB.28.4272} {\bibfield  {journal} {\bibinfo
  {journal} {Phys. Rev. B}\ }\textbf {\bibinfo {volume} {28}},\ \bibinfo
  {pages} {4272} (\bibinfo {year} {1983})}\BibitemShut {NoStop}%
\bibitem [{\citenamefont {Ostlund}\ \emph {et~al.}(1983)\citenamefont
  {Ostlund}, \citenamefont {Pandit}, \citenamefont {Rand}, \citenamefont
  {Schellnhuber},\ and\ \citenamefont {Siggia}}]{Ostlund1983}%
  \BibitemOpen
  \bibfield  {author} {\bibinfo {author} {\bibfnamefont {S.}~\bibnamefont
  {Ostlund}}, \bibinfo {author} {\bibfnamefont {R.}~\bibnamefont {Pandit}},
  \bibinfo {author} {\bibfnamefont {D.}~\bibnamefont {Rand}}, \bibinfo {author}
  {\bibfnamefont {H.~J.}\ \bibnamefont {Schellnhuber}},\ and\ \bibinfo {author}
  {\bibfnamefont {E.~D.}\ \bibnamefont {Siggia}},\ }\bibfield  {title}
  {\bibinfo {title} {{One-dimensional schr{\"{o}}dinger equation with an almost
  periodic potential}},\ }\href {https://doi.org/10.1103/PhysRevLett.50.1873}
  {\bibfield  {journal} {\bibinfo  {journal} {Phys. Rev. Lett.}\ }\textbf
  {\bibinfo {volume} {50}},\ \bibinfo {pages} {1873} (\bibinfo {year}
  {1983})}\BibitemShut {NoStop}%
\bibitem [{\citenamefont {Hiramoto}\ and\ \citenamefont
  {Kohmoto}(1989)}]{Hiramoto1989}%
  \BibitemOpen
  \bibfield  {author} {\bibinfo {author} {\bibfnamefont {H.}~\bibnamefont
  {Hiramoto}}\ and\ \bibinfo {author} {\bibfnamefont {M.}~\bibnamefont
  {Kohmoto}},\ }\bibfield  {title} {\bibinfo {title} {{New localization in a
  quasiperiodic system}},\ }\href {https://doi.org/10.1103/PhysRevLett.62.2714}
  {\bibfield  {journal} {\bibinfo  {journal} {Phys. Rev. Lett.}\ }\textbf
  {\bibinfo {volume} {62}},\ \bibinfo {pages} {2714} (\bibinfo {year}
  {1989})}\BibitemShut {NoStop}%
\bibitem [{\citenamefont {Roati}\ \emph {et~al.}(2008)\citenamefont {Roati},
  \citenamefont {D'Errico}, \citenamefont {Fallani}, \citenamefont {Fattori},
  \citenamefont {Fort}, \citenamefont {Zaccanti}, \citenamefont {Modugno},
  \citenamefont {Modugno},\ and\ \citenamefont {Inguscio}}]{Roati2008}%
  \BibitemOpen
  \bibfield  {author} {\bibinfo {author} {\bibfnamefont {G.}~\bibnamefont
  {Roati}}, \bibinfo {author} {\bibfnamefont {C.}~\bibnamefont {D'Errico}},
  \bibinfo {author} {\bibfnamefont {L.}~\bibnamefont {Fallani}}, \bibinfo
  {author} {\bibfnamefont {M.}~\bibnamefont {Fattori}}, \bibinfo {author}
  {\bibfnamefont {C.}~\bibnamefont {Fort}}, \bibinfo {author} {\bibfnamefont
  {M.}~\bibnamefont {Zaccanti}}, \bibinfo {author} {\bibfnamefont
  {G.}~\bibnamefont {Modugno}}, \bibinfo {author} {\bibfnamefont
  {M.}~\bibnamefont {Modugno}},\ and\ \bibinfo {author} {\bibfnamefont
  {M.}~\bibnamefont {Inguscio}},\ }\bibfield  {title} {\bibinfo {title}
  {{Anderson localization of a non-interacting Bose-Einstein condensate}},\
  }\href {https://doi.org/10.1038/nature07071} {\bibfield  {journal} {\bibinfo
  {journal} {Nature}\ }\textbf {\bibinfo {volume} {453}},\ \bibinfo {pages}
  {895} (\bibinfo {year} {2008})}\BibitemShut {NoStop}%
\bibitem [{\citenamefont {Lahini}\ \emph {et~al.}(2009)\citenamefont {Lahini},
  \citenamefont {Pugatch}, \citenamefont {Pozzi}, \citenamefont {Sorel},
  \citenamefont {Morandotti}, \citenamefont {Davidson},\ and\ \citenamefont
  {Silberberg}}]{Lahini2009}%
  \BibitemOpen
  \bibfield  {author} {\bibinfo {author} {\bibfnamefont {Y.}~\bibnamefont
  {Lahini}}, \bibinfo {author} {\bibfnamefont {R.}~\bibnamefont {Pugatch}},
  \bibinfo {author} {\bibfnamefont {F.}~\bibnamefont {Pozzi}}, \bibinfo
  {author} {\bibfnamefont {M.}~\bibnamefont {Sorel}}, \bibinfo {author}
  {\bibfnamefont {R.}~\bibnamefont {Morandotti}}, \bibinfo {author}
  {\bibfnamefont {N.}~\bibnamefont {Davidson}},\ and\ \bibinfo {author}
  {\bibfnamefont {Y.}~\bibnamefont {Silberberg}},\ }\bibfield  {title}
  {\bibinfo {title} {{Observation of a localization transition in quasiperiodic
  photonic lattices}},\ }\href {https://doi.org/10.1103/PhysRevLett.103.013901}
  {\bibfield  {journal} {\bibinfo  {journal} {Phys. Rev. Lett.}\ }\textbf
  {\bibinfo {volume} {103}},\ \bibinfo {pages} {1} (\bibinfo {year}
  {2009})}\BibitemShut {NoStop}%
\bibitem [{\citenamefont {Kraus}\ \emph {et~al.}(2012)\citenamefont {Kraus},
  \citenamefont {Lahini}, \citenamefont {Ringel}, \citenamefont {Verbin},\ and\
  \citenamefont {Zilberberg}}]{Kraus2012}%
  \BibitemOpen
  \bibfield  {author} {\bibinfo {author} {\bibfnamefont {Y.~E.}\ \bibnamefont
  {Kraus}}, \bibinfo {author} {\bibfnamefont {Y.}~\bibnamefont {Lahini}},
  \bibinfo {author} {\bibfnamefont {Z.}~\bibnamefont {Ringel}}, \bibinfo
  {author} {\bibfnamefont {M.}~\bibnamefont {Verbin}},\ and\ \bibinfo {author}
  {\bibfnamefont {O.}~\bibnamefont {Zilberberg}},\ }\bibfield  {title}
  {\bibinfo {title} {{Topological States and Adiabatic Pumping in
  Quasicrystals}},\ }\href {https://doi.org/10.1103/PhysRevLett.109.106402}
  {\bibfield  {journal} {\bibinfo  {journal} {Phys. Rev. Lett.}\ }\textbf
  {\bibinfo {volume} {109}},\ \bibinfo {pages} {106402} (\bibinfo {year}
  {2012})}\BibitemShut {NoStop}%
\bibitem [{\citenamefont {Schreiber}\ \emph {et~al.}(2015)\citenamefont
  {Schreiber}, \citenamefont {Hodgman}, \citenamefont {Bordia}, \citenamefont
  {L{\"{u}}schen}, \citenamefont {Fischer}, \citenamefont {Vosk}, \citenamefont
  {Altman}, \citenamefont {Schneider},\ and\ \citenamefont
  {Bloch}}]{Schreiber2015}%
  \BibitemOpen
  \bibfield  {author} {\bibinfo {author} {\bibfnamefont {M.}~\bibnamefont
  {Schreiber}}, \bibinfo {author} {\bibfnamefont {S.~S.}\ \bibnamefont
  {Hodgman}}, \bibinfo {author} {\bibfnamefont {P.}~\bibnamefont {Bordia}},
  \bibinfo {author} {\bibfnamefont {H.~P.}\ \bibnamefont {L{\"{u}}schen}},
  \bibinfo {author} {\bibfnamefont {M.~H.}\ \bibnamefont {Fischer}}, \bibinfo
  {author} {\bibfnamefont {R.}~\bibnamefont {Vosk}}, \bibinfo {author}
  {\bibfnamefont {E.}~\bibnamefont {Altman}}, \bibinfo {author} {\bibfnamefont
  {U.}~\bibnamefont {Schneider}},\ and\ \bibinfo {author} {\bibfnamefont
  {I.}~\bibnamefont {Bloch}},\ }\bibfield  {title} {\bibinfo {title}
  {{Observation of many-body localization of interacting fermions in a
  quasirandom optical lattice}},\ }\href
  {https://doi.org/10.1126/science.aaa7432} {\bibfield  {journal} {\bibinfo
  {journal} {Science}\ }\textbf {\bibinfo {volume} {349}},\ \bibinfo {pages}
  {842} (\bibinfo {year} {2015})}\BibitemShut {NoStop}%
\bibitem [{\citenamefont {Bordia}\ \emph {et~al.}(2017)\citenamefont {Bordia},
  \citenamefont {L{\"{u}}schen}, \citenamefont {Schneider}, \citenamefont
  {Knap},\ and\ \citenamefont {Bloch}}]{Bordia2017}%
  \BibitemOpen
  \bibfield  {author} {\bibinfo {author} {\bibfnamefont {P.}~\bibnamefont
  {Bordia}}, \bibinfo {author} {\bibfnamefont {H.}~\bibnamefont
  {L{\"{u}}schen}}, \bibinfo {author} {\bibfnamefont {U.}~\bibnamefont
  {Schneider}}, \bibinfo {author} {\bibfnamefont {M.}~\bibnamefont {Knap}},\
  and\ \bibinfo {author} {\bibfnamefont {I.}~\bibnamefont {Bloch}},\ }\bibfield
   {title} {\bibinfo {title} {{Periodically driving a many-body localized
  quantum system}},\ }\href {https://doi.org/10.1038/nphys4020} {\bibfield
  {journal} {\bibinfo  {journal} {Nat. Phys.}\ }\textbf {\bibinfo {volume}
  {13}},\ \bibinfo {pages} {460} (\bibinfo {year} {2017})}\BibitemShut
  {NoStop}%
\bibitem [{\citenamefont {Hatsugai}\ and\ \citenamefont
  {Kohmoto}(1990)}]{Hatsugai1990}%
  \BibitemOpen
  \bibfield  {author} {\bibinfo {author} {\bibfnamefont {Y.}~\bibnamefont
  {Hatsugai}}\ and\ \bibinfo {author} {\bibfnamefont {M.}~\bibnamefont
  {Kohmoto}},\ }\bibfield  {title} {\bibinfo {title} {{Energy spectrum and the
  quantum Hall effect on the square lattice with next-nearest-neighbor
  hopping}},\ }\href {https://doi.org/10.1103/PhysRevB.42.8282} {\bibfield
  {journal} {\bibinfo  {journal} {Phys. Rev. B}\ }\textbf {\bibinfo {volume}
  {42}},\ \bibinfo {pages} {8282} (\bibinfo {year} {1990})}\BibitemShut
  {NoStop}%
\bibitem [{\citenamefont {Han}\ \emph {et~al.}(1994)\citenamefont {Han},
  \citenamefont {Thouless}, \citenamefont {Hiramoto},\ and\ \citenamefont
  {Kohmoto}}]{Han1994}%
  \BibitemOpen
  \bibfield  {author} {\bibinfo {author} {\bibfnamefont {J.~H.}\ \bibnamefont
  {Han}}, \bibinfo {author} {\bibfnamefont {D.~J.}\ \bibnamefont {Thouless}},
  \bibinfo {author} {\bibfnamefont {H.}~\bibnamefont {Hiramoto}},\ and\
  \bibinfo {author} {\bibfnamefont {M.}~\bibnamefont {Kohmoto}},\ }\bibfield
  {title} {\bibinfo {title} {{Critical and bicritical properties of Harper's
  equation with next-nearest-neighbor coupling}},\ }\href
  {https://doi.org/10.1103/PhysRevB.50.11365} {\bibfield  {journal} {\bibinfo
  {journal} {Phys. Rev. B}\ }\textbf {\bibinfo {volume} {50}},\ \bibinfo
  {pages} {11365} (\bibinfo {year} {1994})}\BibitemShut {NoStop}%
\bibitem [{\citenamefont {Chang}\ \emph {et~al.}(1997)\citenamefont {Chang},
  \citenamefont {Ikezawa},\ and\ \citenamefont {Kohmoto}}]{Chang1997}%
  \BibitemOpen
  \bibfield  {author} {\bibinfo {author} {\bibfnamefont {I.}~\bibnamefont
  {Chang}}, \bibinfo {author} {\bibfnamefont {K.}~\bibnamefont {Ikezawa}},\
  and\ \bibinfo {author} {\bibfnamefont {M.}~\bibnamefont {Kohmoto}},\
  }\bibfield  {title} {\bibinfo {title} {{Multifractal properties of the wave
  functions of the square-lattice tight-binding model with
  next-nearest-neighbor hopping in a magnetic field}},\ }\href
  {https://doi.org/10.1103/PhysRevB.55.12971} {\bibfield  {journal} {\bibinfo
  {journal} {Phys. Rev. B}\ }\textbf {\bibinfo {volume} {55}},\ \bibinfo
  {pages} {12971} (\bibinfo {year} {1997})}\BibitemShut {NoStop}%
\bibitem [{\citenamefont {Takada}\ \emph {et~al.}(2004)\citenamefont {Takada},
  \citenamefont {Ino},\ and\ \citenamefont {Yamanaka}}]{Takada2004}%
  \BibitemOpen
  \bibfield  {author} {\bibinfo {author} {\bibfnamefont {Y.}~\bibnamefont
  {Takada}}, \bibinfo {author} {\bibfnamefont {K.}~\bibnamefont {Ino}},\ and\
  \bibinfo {author} {\bibfnamefont {M.}~\bibnamefont {Yamanaka}},\ }\bibfield
  {title} {\bibinfo {title} {{Statistics of spectra for critical quantum chaos
  in one-dimensional quasiperiodic systems}},\ }\href
  {https://doi.org/10.1103/PhysRevE.70.066203} {\bibfield  {journal} {\bibinfo
  {journal} {Phys. Rev. E}\ }\textbf {\bibinfo {volume} {70}},\ \bibinfo
  {pages} {066203} (\bibinfo {year} {2004})}\BibitemShut {NoStop}%
\bibitem [{\citenamefont {Gong}\ and\ \citenamefont {Tong}(2008)}]{Gong2008}%
  \BibitemOpen
  \bibfield  {author} {\bibinfo {author} {\bibfnamefont {L.}~\bibnamefont
  {Gong}}\ and\ \bibinfo {author} {\bibfnamefont {P.}~\bibnamefont {Tong}},\
  }\bibfield  {title} {\bibinfo {title} {{Fidelity, fidelity susceptibility,
  and von Neumann entropy to characterize the phase diagram of an extended
  Harper model}},\ }\href {https://doi.org/10.1103/PhysRevB.78.115114}
  {\bibfield  {journal} {\bibinfo  {journal} {Phys. Rev. B - Condens. Matter
  Mater. Phys.}\ }\textbf {\bibinfo {volume} {78}},\ \bibinfo {pages} {1}
  (\bibinfo {year} {2008})}\BibitemShut {NoStop}%
\bibitem [{\citenamefont {Liu}\ \emph {et~al.}(2015)\citenamefont {Liu},
  \citenamefont {Ghosh},\ and\ \citenamefont {Chong}}]{Liu2015}%
  \BibitemOpen
  \bibfield  {author} {\bibinfo {author} {\bibfnamefont {F.}~\bibnamefont
  {Liu}}, \bibinfo {author} {\bibfnamefont {S.}~\bibnamefont {Ghosh}},\ and\
  \bibinfo {author} {\bibfnamefont {Y.~D.}\ \bibnamefont {Chong}},\ }\bibfield
  {title} {\bibinfo {title} {{Localization and adiabatic pumping in a
  generalized Aubry-Andr{\'{e}}-Harper model}},\ }\href
  {https://doi.org/10.1103/PhysRevB.91.014108} {\bibfield  {journal} {\bibinfo
  {journal} {Phys. Rev. B}\ }\textbf {\bibinfo {volume} {91}},\ \bibinfo
  {pages} {014108} (\bibinfo {year} {2015})}\BibitemShut {NoStop}%
\bibitem [{\citenamefont {Zhao}\ \emph {et~al.}(2017)\citenamefont {Zhao},
  \citenamefont {Shi}, \citenamefont {Yu},\ and\ \citenamefont
  {Yi}}]{Zhao2017}%
  \BibitemOpen
  \bibfield  {author} {\bibinfo {author} {\bibfnamefont {X.~L.}\ \bibnamefont
  {Zhao}}, \bibinfo {author} {\bibfnamefont {Z.~C.}\ \bibnamefont {Shi}},
  \bibinfo {author} {\bibfnamefont {C.~S.}\ \bibnamefont {Yu}},\ and\ \bibinfo
  {author} {\bibfnamefont {X.~X.}\ \bibnamefont {Yi}},\ }\bibfield  {title}
  {\bibinfo {title} {{Influence of localization transition on dynamical
  properties for an extended Aubry–Andr{\'{e}}–Harper model}},\ }\href
  {https://doi.org/10.1088/1361-6455/aa92df} {\bibfield  {journal} {\bibinfo
  {journal} {J. Phys. B At. Mol. Opt. Phys.}\ }\textbf {\bibinfo {volume}
  {50}},\ \bibinfo {pages} {235503} (\bibinfo {year} {2017})}\BibitemShut
  {NoStop}%
\bibitem [{\citenamefont {Wang}\ \emph {et~al.}(2021)\citenamefont {Wang},
  \citenamefont {Cheng}, \citenamefont {Liu},\ and\ \citenamefont
  {Yu}}]{Wang2021}%
  \BibitemOpen
  \bibfield  {author} {\bibinfo {author} {\bibfnamefont {Y.}~\bibnamefont
  {Wang}}, \bibinfo {author} {\bibfnamefont {C.}~\bibnamefont {Cheng}},
  \bibinfo {author} {\bibfnamefont {X.-J.}\ \bibnamefont {Liu}},\ and\ \bibinfo
  {author} {\bibfnamefont {D.}~\bibnamefont {Yu}},\ }\bibfield  {title}
  {\bibinfo {title} {{Many-Body Critical Phase: Extended and Nonthermal}},\
  }\href {https://doi.org/10.1103/PhysRevLett.126.080602} {\bibfield  {journal}
  {\bibinfo  {journal} {Phys. Rev. Lett.}\ }\textbf {\bibinfo {volume} {126}},\
  \bibinfo {pages} {080602} (\bibinfo {year} {2021})}\BibitemShut {NoStop}%
\bibitem [{\citenamefont {Xiao}\ \emph {et~al.}(2021)\citenamefont {Xiao},
  \citenamefont {Xie}, \citenamefont {Dong}, \citenamefont {Chen},
  \citenamefont {Yi},\ and\ \citenamefont {Yan}}]{Xiao2021}%
  \BibitemOpen
  \bibfield  {author} {\bibinfo {author} {\bibfnamefont {T.}~\bibnamefont
  {Xiao}}, \bibinfo {author} {\bibfnamefont {D.}~\bibnamefont {Xie}}, \bibinfo
  {author} {\bibfnamefont {Z.}~\bibnamefont {Dong}}, \bibinfo {author}
  {\bibfnamefont {T.}~\bibnamefont {Chen}}, \bibinfo {author} {\bibfnamefont
  {W.}~\bibnamefont {Yi}},\ and\ \bibinfo {author} {\bibfnamefont
  {B.}~\bibnamefont {Yan}},\ }\bibfield  {title} {\bibinfo {title}
  {{Observation of topological phase with critical localization in a
  quasi-periodic lattice}},\ }\href
  {https://doi.org/10.1016/j.scib.2021.07.025} {\bibfield  {journal} {\bibinfo
  {journal} {Sci. Bull.}\ }\textbf {\bibinfo {volume} {66}},\ \bibinfo {pages}
  {2175} (\bibinfo {year} {2021})}\BibitemShut {NoStop}%
\bibitem [{\citenamefont {Yan}\ \emph {et~al.}(2018)\citenamefont {Yan},
  \citenamefont {Krantz}, \citenamefont {Sung}, \citenamefont {Kjaergaard},
  \citenamefont {Campbell}, \citenamefont {Orlando}, \citenamefont
  {Gustavsson},\ and\ \citenamefont {Oliver}}]{Yan2018}%
  \BibitemOpen
  \bibfield  {author} {\bibinfo {author} {\bibfnamefont {F.}~\bibnamefont
  {Yan}}, \bibinfo {author} {\bibfnamefont {P.}~\bibnamefont {Krantz}},
  \bibinfo {author} {\bibfnamefont {Y.}~\bibnamefont {Sung}}, \bibinfo {author}
  {\bibfnamefont {M.}~\bibnamefont {Kjaergaard}}, \bibinfo {author}
  {\bibfnamefont {D.~L.}\ \bibnamefont {Campbell}}, \bibinfo {author}
  {\bibfnamefont {T.~P.}\ \bibnamefont {Orlando}}, \bibinfo {author}
  {\bibfnamefont {S.}~\bibnamefont {Gustavsson}},\ and\ \bibinfo {author}
  {\bibfnamefont {W.~D.}\ \bibnamefont {Oliver}},\ }\bibfield  {title}
  {\bibinfo {title} {{Tunable Coupling Scheme for Implementing High-Fidelity
  Two-Qubit Gates}},\ }\href {https://doi.org/10.1103/PhysRevApplied.10.054062}
  {\bibfield  {journal} {\bibinfo  {journal} {Phys. Rev. Appl.}\ }\textbf
  {\bibinfo {volume} {10}},\ \bibinfo {pages} {1} (\bibinfo {year}
  {2018})}\BibitemShut {NoStop}%
\bibitem [{\citenamefont {Shi}\ \emph {et~al.}()\citenamefont {Shi},
  \citenamefont {Yang}, \citenamefont {Xiang}, \citenamefont {Ge},
  \citenamefont {Li}, \citenamefont {Wang}, \citenamefont {Huang},
  \citenamefont {Tian}, \citenamefont {Song}, \citenamefont {Zheng},
  \citenamefont {Xu}, \citenamefont {Cai},\ and\ \citenamefont
  {Fan}}]{Shi2021}%
  \BibitemOpen
  \bibfield  {author} {\bibinfo {author} {\bibfnamefont {Y.-H.}\ \bibnamefont
  {Shi}}, \bibinfo {author} {\bibfnamefont {R.-Q.}\ \bibnamefont {Yang}},
  \bibinfo {author} {\bibfnamefont {Z.}~\bibnamefont {Xiang}}, \bibinfo
  {author} {\bibfnamefont {Z.-Y.}\ \bibnamefont {Ge}}, \bibinfo {author}
  {\bibfnamefont {H.}~\bibnamefont {Li}}, \bibinfo {author} {\bibfnamefont
  {Y.-Y.}\ \bibnamefont {Wang}}, \bibinfo {author} {\bibfnamefont
  {K.}~\bibnamefont {Huang}}, \bibinfo {author} {\bibfnamefont
  {Y.}~\bibnamefont {Tian}}, \bibinfo {author} {\bibfnamefont {X.}~\bibnamefont
  {Song}}, \bibinfo {author} {\bibfnamefont {D.}~\bibnamefont {Zheng}},
  \bibinfo {author} {\bibfnamefont {K.}~\bibnamefont {Xu}}, \bibinfo {author}
  {\bibfnamefont {R.-G.}\ \bibnamefont {Cai}},\ and\ \bibinfo {author}
  {\bibfnamefont {H.}~\bibnamefont {Fan}},\ }\bibfield  {title} {\bibinfo
  {title} {{On-chip black hole: Hawking radiation and curved spacetime in a
  superconducting quantum circuit with tunable couplers}},\ }\href
  {http://arxiv.org/abs/2111.11092} {\ }\Eprint
  {https://arxiv.org/abs/2111.11092} {arXiv:2111.11092} \BibitemShut {NoStop}%
\bibitem [{\citenamefont {Thouless}(1974)}]{THOULESS197493}%
  \BibitemOpen
  \bibfield  {author} {\bibinfo {author} {\bibfnamefont {D.~J.}\ \bibnamefont
  {Thouless}},\ }\bibfield  {title} {\bibinfo {title} {{Electrons in disordered
  systems and the theory of localization}},\ }\href
  {https://doi.org/https://doi.org/10.1016/0370-1573(74)90029-5} {\bibfield
  {journal} {\bibinfo  {journal} {Phys. Rep.}\ }\textbf {\bibinfo {volume}
  {13}},\ \bibinfo {pages} {93} (\bibinfo {year} {1974})}\BibitemShut {NoStop}%
\bibitem [{\citenamefont {Bell}(1972)}]{Bell1972}%
  \BibitemOpen
  \bibfield  {author} {\bibinfo {author} {\bibfnamefont {R.~J.}\ \bibnamefont
  {Bell}},\ }\bibfield  {title} {\bibinfo {title} {{The dynamics of disordered
  lattices}},\ }\href {https://doi.org/10.1088/0034-4885/35/3/306} {\bibfield
  {journal} {\bibinfo  {journal} {Reports Prog. Phys.}\ }\textbf {\bibinfo
  {volume} {35}},\ \bibinfo {pages} {306} (\bibinfo {year} {1972})}\BibitemShut
  {NoStop}%
\bibitem [{\citenamefont {Sch\"afer}\ and\ \citenamefont
  {Wegner}(1980)}]{Schfer1980}%
  \BibitemOpen
  \bibfield  {author} {\bibinfo {author} {\bibfnamefont {L.}~\bibnamefont
  {Sch\"afer}}\ and\ \bibinfo {author} {\bibfnamefont {F.}~\bibnamefont
  {Wegner}},\ }\bibfield  {title} {\bibinfo {title} {{Lattice instantons, a
  basis for a treatment of localized states?}},\ }\href
  {https://doi.org/10.1007/BF01305826} {\bibfield  {journal} {\bibinfo
  {journal} {Zeitschrift f\"ur Phys. B Condens. Matter}\ }\textbf {\bibinfo
  {volume} {39}},\ \bibinfo {pages} {281} (\bibinfo {year} {1980})}\BibitemShut
  {NoStop}%
\bibitem [{\citenamefont {Rodriguez}\ \emph {et~al.}(2011)\citenamefont
  {Rodriguez}, \citenamefont {Vasquez}, \citenamefont {Slevin},\ and\
  \citenamefont {R{\"{o}}mer}}]{Rodriguez2011}%
  \BibitemOpen
  \bibfield  {author} {\bibinfo {author} {\bibfnamefont {A.}~\bibnamefont
  {Rodriguez}}, \bibinfo {author} {\bibfnamefont {L.~J.}\ \bibnamefont
  {Vasquez}}, \bibinfo {author} {\bibfnamefont {K.}~\bibnamefont {Slevin}},\
  and\ \bibinfo {author} {\bibfnamefont {R.~A.}\ \bibnamefont {R{\"{o}}mer}},\
  }\bibfield  {title} {\bibinfo {title} {{Multifractal finite-size scaling and
  universality at the Anderson transition}},\ }\href
  {https://doi.org/10.1103/PhysRevB.84.134209} {\bibfield  {journal} {\bibinfo
  {journal} {Phys. Rev. B}\ }\textbf {\bibinfo {volume} {84}},\ \bibinfo
  {pages} {134209} (\bibinfo {year} {2011})}\BibitemShut {NoStop}%
\bibitem [{\citenamefont {Luitz}\ \emph {et~al.}(2014)\citenamefont {Luitz},
  \citenamefont {Alet},\ and\ \citenamefont {Laflorencie}}]{Luitz2014}%
  \BibitemOpen
  \bibfield  {author} {\bibinfo {author} {\bibfnamefont {D.~J.}\ \bibnamefont
  {Luitz}}, \bibinfo {author} {\bibfnamefont {F.}~\bibnamefont {Alet}},\ and\
  \bibinfo {author} {\bibfnamefont {N.}~\bibnamefont {Laflorencie}},\
  }\bibfield  {title} {\bibinfo {title} {{Universal Behavior beyond
  Multifractality in Quantum Many-Body Systems}},\ }\href
  {https://doi.org/10.1103/PhysRevLett.112.057203} {\bibfield  {journal}
  {\bibinfo  {journal} {Phys. Rev. Lett.}\ }\textbf {\bibinfo {volume} {112}},\
  \bibinfo {pages} {057203} (\bibinfo {year} {2014})}\BibitemShut {NoStop}%
\bibitem [{\citenamefont {Luitz}\ \emph {et~al.}(2015)\citenamefont {Luitz},
  \citenamefont {Laflorencie},\ and\ \citenamefont {Alet}}]{Luitz2015}%
  \BibitemOpen
  \bibfield  {author} {\bibinfo {author} {\bibfnamefont {D.~J.}\ \bibnamefont
  {Luitz}}, \bibinfo {author} {\bibfnamefont {N.}~\bibnamefont {Laflorencie}},\
  and\ \bibinfo {author} {\bibfnamefont {F.}~\bibnamefont {Alet}},\ }\bibfield
  {title} {\bibinfo {title} {{Many-body localization edge in the random-field
  Heisenberg chain}},\ }\href {https://doi.org/10.1103/PhysRevB.91.081103}
  {\bibfield  {journal} {\bibinfo  {journal} {Phys. Rev. B}\ }\textbf {\bibinfo
  {volume} {91}},\ \bibinfo {pages} {081103} (\bibinfo {year}
  {2015})}\BibitemShut {NoStop}%
\end{thebibliography}%
\clearpage

\onecolumngrid
\begin{center}
{\large \bf Supplementary Materials:
 \\ Observation of critical phase transition in a generalized Aubry-Andr\'e-Harper model on a superconducting quantum processor }\\
\vspace{0.3cm}
\end{center}

\vspace{0.6cm}
\twocolumngrid

\beginsupplement
\section{Qubit Information}
Our experiment is performed on the superconducting quantum processor with tunable couplers, which is identical to the one in reference \cite{Shi2021}. This device consists of 10 transmon qubits ($Q_1 \sim Q_{10}$) and 9 transmon-type couplers ($C_1 \sim C_9$). In this experiment all qubits are initilized at their idle frequencies $\omega_{\mathrm{idle}}^j/2\pi$ ($j=1 , 2, \cdots, 10$) that spread in the range from 4.280 GHz to 4.900 GHz. The idle frequencies are carefully arranged to reduce unwanted interaction and  crosstalk errors among qubits (or between qubits and couplers) during single-qubit operations. All relevant information about qubit characteristics are listed in Table~\ref{tab1}. The anharmonicity of all qubits are around 200 MHz.  The readout pulse for all qubits are $1.0 \mu s$ in length, and the readout frequency and readout power are optimized for a high visibility. The readout fidelity, which is denoted by $F_0$ and $F_1$, are listed in Table~\ref{tab1}.
\begin{table*}[ht]
	\centering
	\caption{Qubit charateristics. $\omega_{\textrm{idle}}$ is the idle frequency of qubit where decoherence parameters (including energy relaxation time $T_{1}$ and Ramsey Gaussian dephasing time $T_{2}^{\ast}$) are measured. $\omega_{\textrm{max}}/2\pi$ and $\omega_{\textrm{r}}/2\pi$ denote to qubits' sweetpoint frequency and resonant frequency, respectively. $F_{0}$ ($F_{1}$) is the measurement probability of $|0\rangle$ ($|1\rangle$) when the qubit is prepared in $|0\rangle$ ($|1\rangle$), which is used to mitigate the readout errors.
	}
	\label{tab1}

	\setlength\tabcolsep{2.5mm}{
	\begin{tabular}{ccccccccccc}
		
		\hline
		\hline
		 &   Q$_{1}$  &   Q$_{2}$  &  Q$_{3}$  &   Q$_{4}$  &   Q$_{5}$  &   Q$_{6}$  
		 &   Q$_{7}$ & Q$_{8}$ & Q$_{9}$  & Q$_{10}$\\
		\hline
		$\omega_{\textrm{idle}}/2\pi$ (GHz) & 4.900 & 4.365 & 4.930 & 4.310 & 4.878 & 4.455 & 4.800 & 4.280 & 4.770 & 4.388\\
		
		$\omega_{\textrm{r}}/2\pi$ (GHz) & 6.857 & 6.687 & 6.667 & 6.733 & 6.708 & 6.772 & 6.749 & 6.815 & 6.792 & 6.839\\
		
		$\omega_{\textrm{max}}/2\pi$ (GHz) & 5.491 & 5.201 & 5.347 & 5.36  & 5.226 & 5.313 & 5.318 & 5.3   & 5.108 & 5.248\\
		
		$T_{1}$ ($\mu s$) & 14.3 & 14.8 & 33.0 & 17.9 & 30.7 & 28.1 & 26.3 & 21.5 & 24.4 & 32.9 \\
		$T_{2}^{*}$ ($\mu s$) & 0.9 & 1.1 & 1.1 & 1.2 & 1.2 & 1.1 & 1.2 & 1.6 & 1.2 & 1.3 \\
		$F_{0}$ & 0.969 & 0.947 & 0.966 & 0.956 & 0.956 & 0.948 & 0.968 & 0.951 & 0.957 & 0.954 \\
		$F_{1}$ & 0.926 & 0.901 & 0.926 & 0.907 & 0.918 & 0.909 & 0.915 & 0.895 & 0.919 & 0.913\\
		\hline
		\hline& & 
	\end{tabular}}
\end{table*}

\section{Z pulse crosstalk correction}
\begin{figure}[ht]
	\begin{center}
		\includegraphics[width=1\linewidth,clip=True]{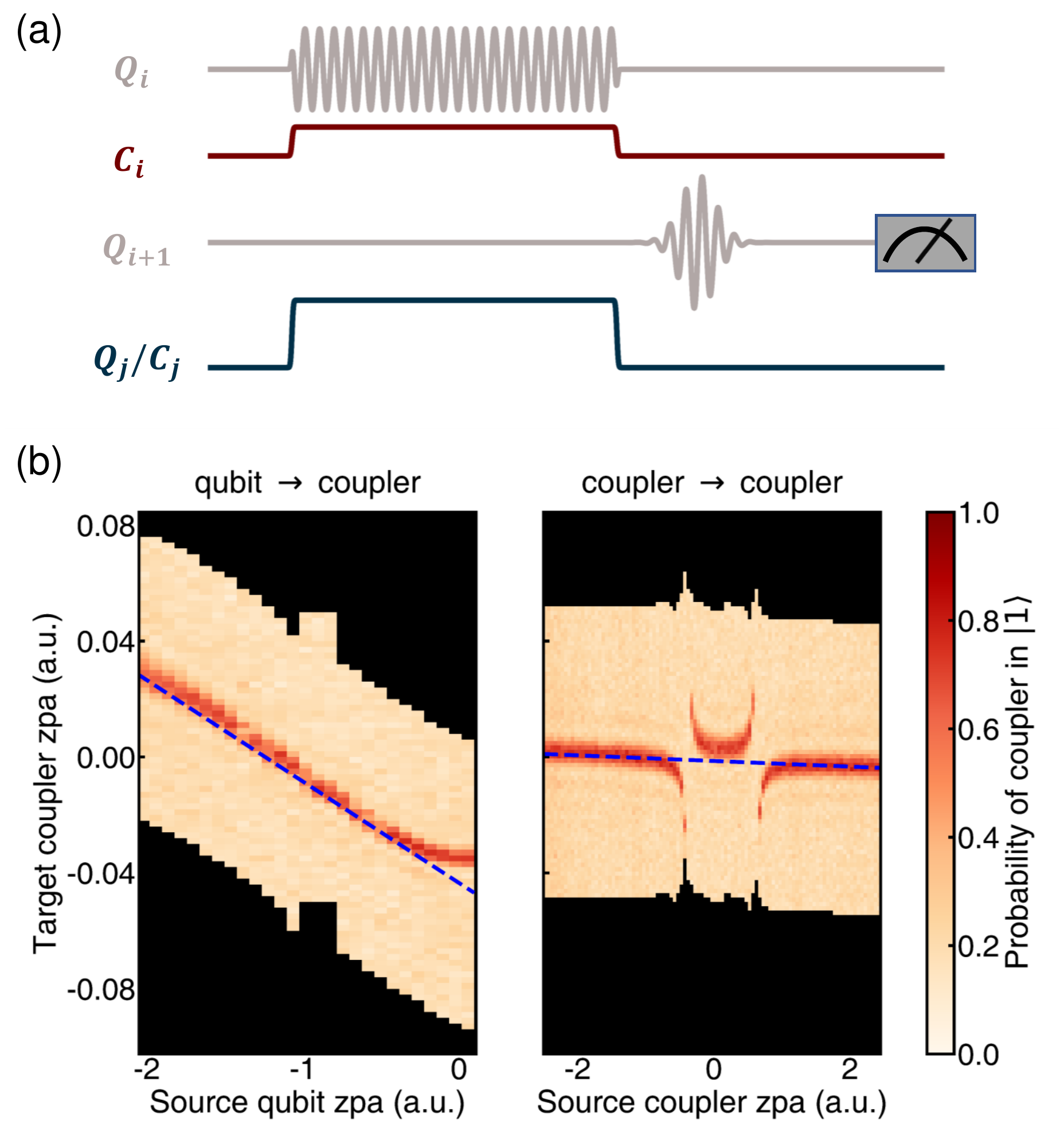}
		\caption{
				Measurement of qubit- and coupler-to-coupler Z crosstalks. (a) Pulse sequence used for measurement. Here $Q_i$ and $Q_j$ are used to excite and measure the target coupler $C_i$, respectively. (b) Typical data of qubit-to-coupler (left) and coupler-to-coupler (right) Z crosstalk. The blue dash lines are the results of linear fitting to obtain the Z crosstalk matrix elements.}
		\label{fig:Zcrosstalk_pulse}
	\end{center}
\end{figure}

\begin{figure}[hb]
	\begin{center}
		\includegraphics[width=1\linewidth,clip=True]{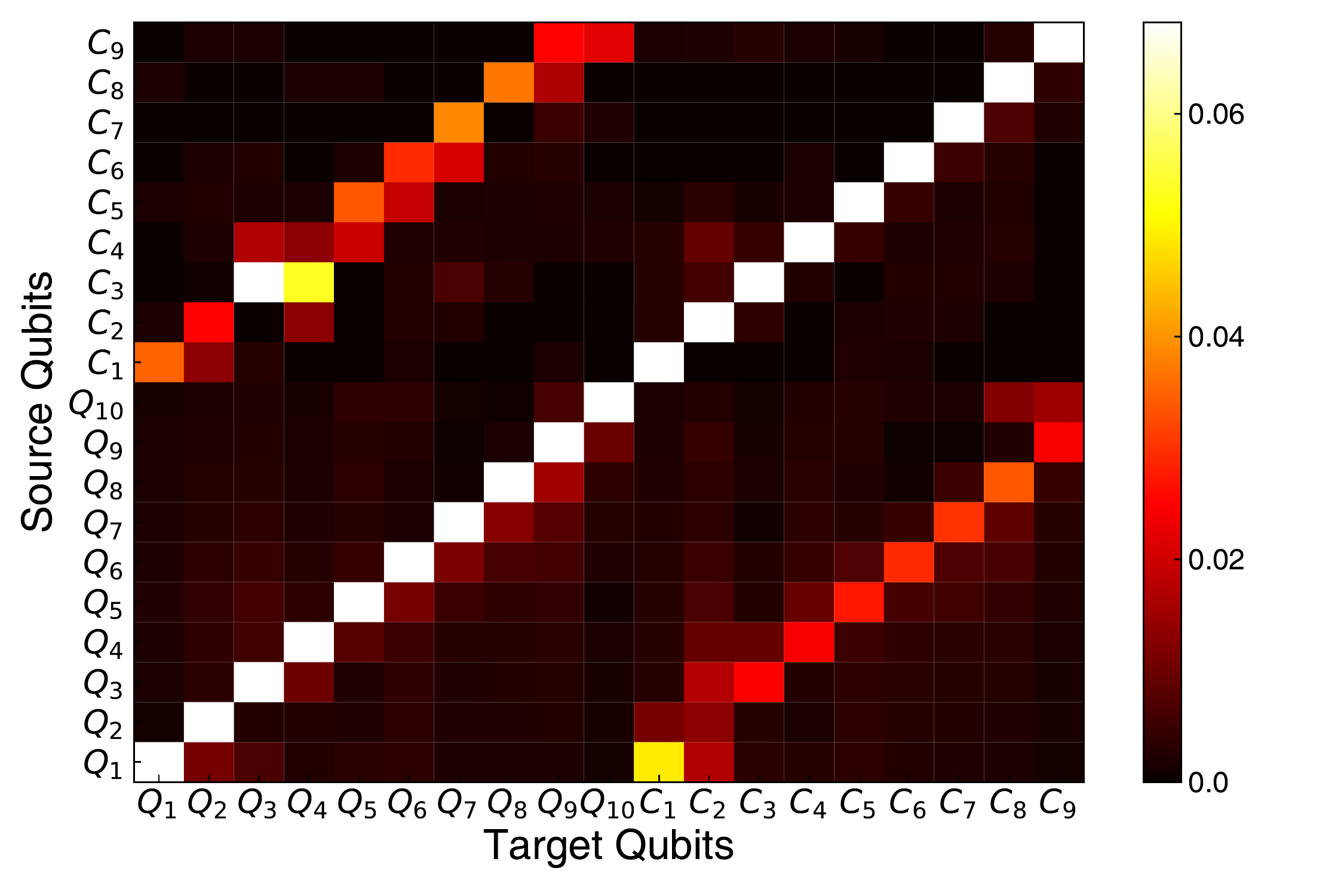}
		\caption{
				Z line crosstalk matrix.
			All the element values between qubit pairs or coupler pairs is at a low level ($< 2 \%$) compared with high crosstalk between qubits and couplers.}
		\label{fig:Zcrosstalk}
	\end{center}
\end{figure}

The Z control line crosstalks are estimated by measuring the linear response with respect to the bias voltages. To achieve this, we excite the target and measure how much Z pulse amplitude (zpa) of the target needs to compensate for the crosstalk from the source. It is routine to measure crosstalk between qubits, but specific means are required to measure crosstalk between qubits and couplers. Following Ref.~\cite{Shi2021}, we measure the coupler-to-qubit Z crosstalk and further add the measurement of qubit- and coupler-to-coupler Z crosstalks. Pulse sequence used for such measurement of crosstalk to couplers is plotted in Fig.~\ref{fig:Zcrosstalk_pulse}(a). Due to the AC Stark effect, the response of target zpa to source zpa is non-linear. In order to obtain the classical Z crosstalk that removes the AC Stark effect, we fit the data in the linear region where the source qubit (coupler) is far from the target coupler, see Fig.~\ref{fig:Zcrosstalk_pulse}(b). We then obtain the total crosstalk matrix, as shown in Fig.~\ref{fig:Zcrosstalk}. In our device, all crosstalks between qubits are reduced due to its coupler architecture. 

\section{Measurement of effective coupling strengths}
The system can be fully described by the generalized Aubry-Andr\'e-Harper (GAAH) model with the Hamiltonian:
\begin{eqnarray}
	&\frac{\hat H }{\hbar} = \sum\limits_{j=1}^{9} ({ J_{j, j+1}{\hat a}_j^{\dagger} {{\hat a}_{j + 1}}+ \text{H.c.}})+\sum\limits_{j=1}^{10} { h_j{\hat a}_j^{\dagger} {{\hat a}_{j}}}.
	\label{eq_SGAAH}
\end{eqnarray}
Here $J_{j, j+1} = \lambda \left( 1 + \mu \cos \left[ {2\pi \left( {j + \frac{1}{2}} \right)\alpha  + \delta } \right]\right)$ and $h_j = \lambda V\cos (2\pi j\alpha  + \delta )$ represent the effective coupling of qubits and the frequency detuning relative to resonant frequency ($\simeq4.36$ GHz). In this paper, we set $\lambda/2\pi=4$ MHz and $\alpha = \left(\sqrt{5}-1\right)/2$. $\mu$ and $V$ represent modulations amplitudes of off-diagonal hopping and on-site incommensurate potential, respectively. We manipulate $\mu$ and $V$ by applying the fast voltages to the corresponding Z control lines of the coupler and qubit respectively.

\begin{figure}[ht]
	\begin{center}
		\includegraphics[width=1\linewidth,clip=True]{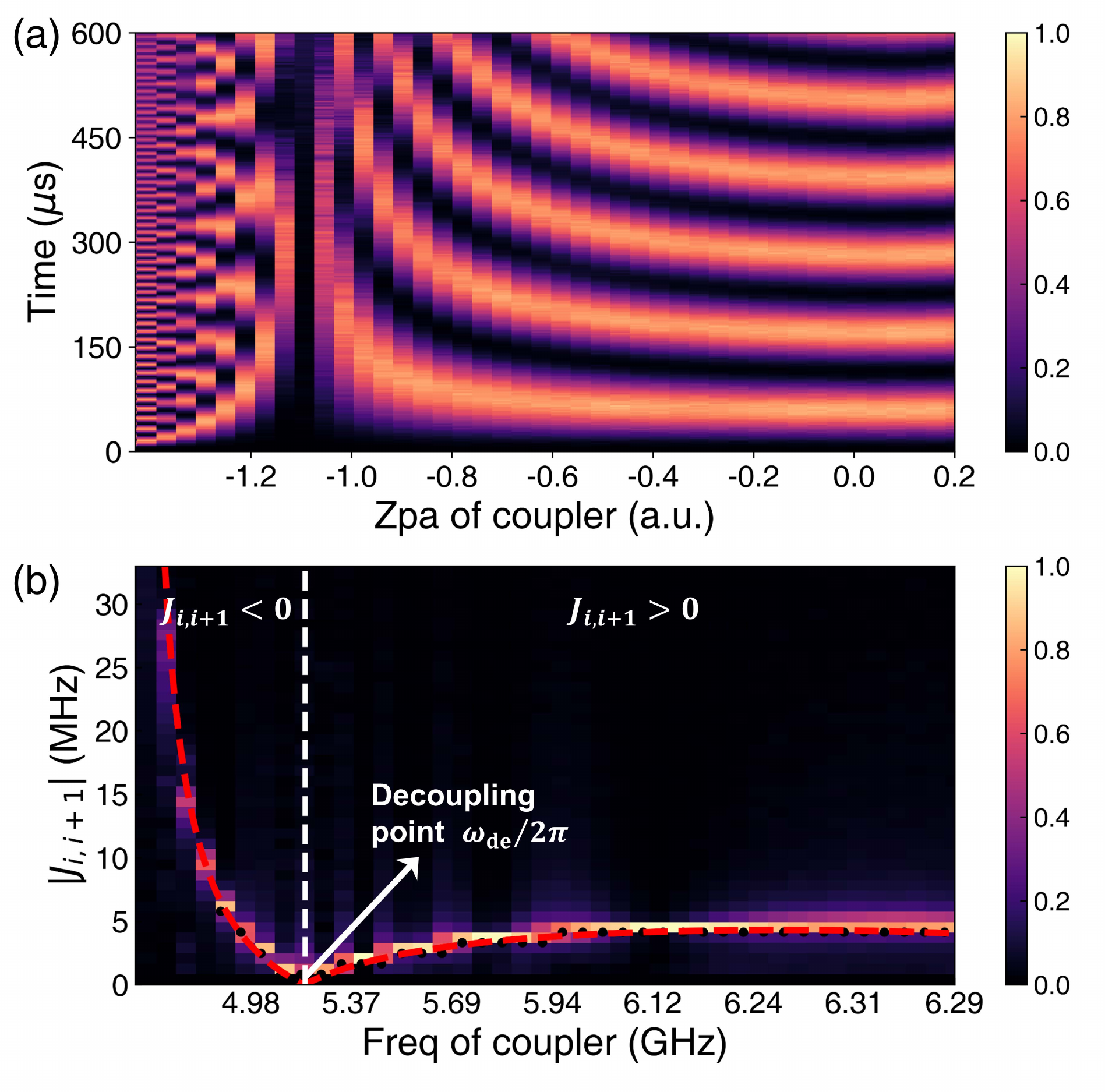}
		\caption{
				Experiment data of effective coupling strength.
			(a) Joint probability $P_{\rm{01}}\left(t\right)$  of qubits with varied coupler zpa. (b) The corresponding normalized Fourier transformation amplitude versus coupler frequency $\omega_{c_i}/2\pi$. The effective coupling strength is calculated as half the Fourier frequency of probability $P_{01}\left(t\right)$. The distribution of ${J}_{i,i+1}$, marked red dashed line, is calculated by fitting the peak value, while the white dashed line denotes one decoupling point $\omega_\mathrm{de}$ (i.e., ${J}_{i,i+1}=0$). When $\omega_{c_i}>\omega_{\mathrm{de}}$, ${J}_{i,i+1}$ decreases from a positive value with $\omega_{c_i}$  until the decoupling point. ${J}_{i,i+1}$ becomes negative in the region $\omega_{c_i}<\omega_{\mathrm{de}}$ and its absolute value increases rapidly which makes large coupling strength possible.}
		\label{fig:swap}
	\end{center}
\end{figure}

In our processor of coupler architecture, the effective coupling of qubits can be adjusted by tuning couplers frequencies. Mathematically, the effective coupling strength can be described by \cite{Yan2018,Shi2021}:

\begin{equation}
	J_{j,j+1} = J_{j,j+1}^{0} + \frac{J_j^0 J_{j+1}^0}{\Delta_{j,j+1}},
	\label{eq_EffCS}
\end{equation}
where $J_{j,j+1}^{0}$ is the direct coupling between $Q_j$ and $Q_{j+1}$, $J_j^0$ is the direct coupling between $Q_j$ and $C_j$, and $1/\Delta_{j,j+1} = \left[1/\left(\omega_{Q_{j}}-\omega_{C_{j}}\right)+1/\left(\omega_{Q_{j+1}}-\omega_{C_{j}}\right)\right]/2$ is related to frequency detuning between the $j$-th coupler and its two nearest neighbor qubits. As shown in Fig.~\ref{fig:swap}(a), effective coupling strengths can be measured precisely through the joint probability as a function of qubit-qubit swapping time $t$ and coupler zpa \cite{Shi2021}. Fig.~\ref{fig:swap}(b) is the normalized Fourier transformation corresponding to Fig.~\ref{fig:swap}(a). Fig.~\ref{fig:swap}(b) shows the advantages of adjusting the effective coupling strength using the coupler architecture. Varying coupler zpa not only makes decoupling point possible, bu also greatly improve the upper limit of coupling strength. In the experiment, we first calculate the coupling strength distribution satisfying Eq.~\eqref{eq_GAAH} according to Eq.~\eqref{eq_EffCS} and estimate the zpa of each coupler, and then fine-tune the zpa by performing two-qubit swapping experiment until the coupling strength is within the tolerance ($\leq0.1$MHz).

\begin{figure}[ht]
	\begin{center}
		\includegraphics[width=1\linewidth,clip=True]{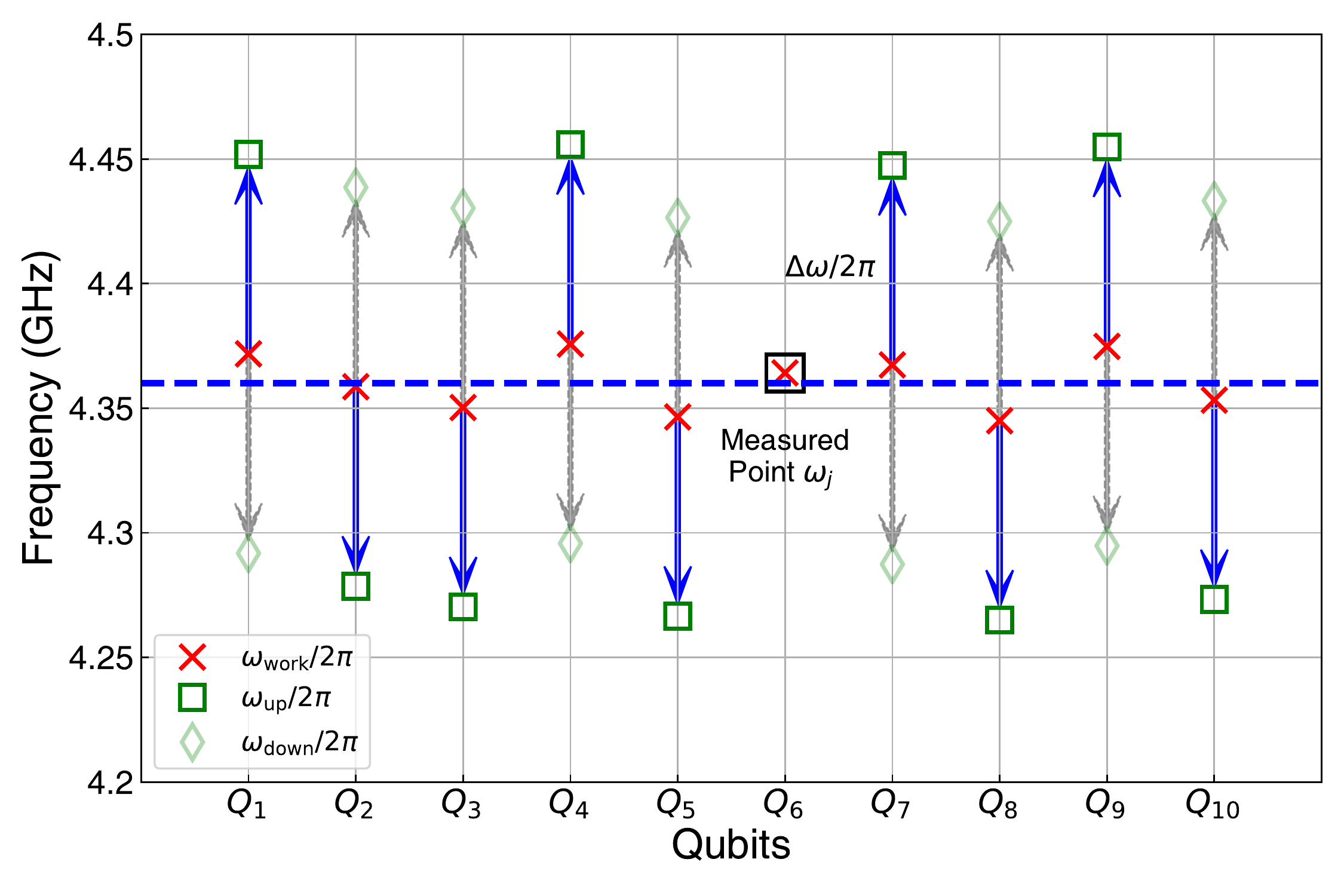}
		\caption{
				Schematic diagram of tune zpa of qubit precisely with $Q_6$ as the target qubit. 
				The red cross represents working frequency $\omega_\mathrm{work}/2\pi$. The squares and diamond represent two frequency distributions. We tune the zpa of $Q_6$ with other qubits shifted by $\Delta \omega/2\pi$ (Here we set $\Delta \omega/2\pi = 80 \rm{MHz}$ in this paper) away from the target qubit (or close to target qubit) based on their own working points.
			}
		\label{fig:tunefreq}
	\end{center}
\end{figure}

After all the coupling strengths are calibrated, we need to correct the zpas that bias qubits frequencies. To eliminate the zpa deviation caused by the AC Stark effect, we perform two specific Rabi oscillation experiments on the target qubit whose zpa needs to be determined. In details, we consider two staggered distributions of non-target qubits frequencies as shown in Fig.~\ref{fig:tunefreq}. Assuming that the zpa of target qubit measured by the first Rabi experiment with one of staggered frequency distribution is $\text{zpa}_1$, and the second Rabi experiment corresponding to the reverse-staggered frequency distribution measures another $\text{zpa}_2$, we can estimate the zpa of target qubit that bias it to working point as $(\text{zpa}_1+\text{zpa}_2)/2$. Thus, the zpa deviation caused by the AC Stark effect that depends on the frequency detuning can be cancelled approximately. We calibrate all the qubits zpas one by one in this way, and finally bias the frequencies of all the qubits to the specified distribution.

\section{Effects of decoherence}
In order to increase the adjustment range of coupling strength and reduce the distortion caused by high zpa, we choose a low resonant frequency ($\simeq4.36$ GHz, deviates about 800 MHz from the sweet point) and decrease the idle frequencies $\omega_{\mathrm{idle}}$ accordingly. However, as a consequence, the decoherence of qubits is exacerbated by low frequency noise, leading to the decrease of dephasing time $T_2^{\ast}$. Besides, the existence of energy relaxation time $T_1$ may also contribute to the dissipation. Such dissipation dynamics can be described by the Lindblad master equation
\begin{equation}
\frac{\mathrm{d}\hat{\rho}(t)}{\mathrm{d}t}=-\frac{\mathrm{i}}{\hbar}\left[\hat{H},\hat{\rho(t)}\right]+\sum_{n=1}^{L}\left(\hat{K}_{n}\hat{\rho}(t) \hat{K}^{\dagger}_n-\frac{1}{2}\left\{\hat{K}^{\dagger}_{n}\hat{K}_n,\hat{\rho}(t)\right\}\right),
\label{eq_Lindblad}
\end{equation}
with $\hat{\rho(t)}$ being the time-dependent density matrix of $L$ qubits and $\hat{K}^{\dagger}_{n}$ ($\hat{K}^{\dagger}_{n}$) as Lindblad operators.
The first term of Eq.~\eqref{eq_Lindblad} represents the unitary evolutions of the system and the other terms imply the dissipation of the system due to interaction with the environment. Here we consider the dephasing and energy relaxation effects, and the corresponding Lindblad operators are written as $\hat{K}_{n}=(1-2\hat{a}^{\dagger}_n\hat{a}_n)/\sqrt{2T_2}$ and $\hat{K}^{\dagger}_{n}=\hat{a}_n/\sqrt{T_1}$, respectively.

The numerical results of the dynamics of participation entropies with decoherence effects, shown in Fig.~\ref{fig2}--\ref{fig3} in the main text, and Fig.~\ref{supp_pe1_t} in Supplementary Material, are obtained by solving Eq.~\eqref{eq_Lindblad} with the averaged $\overline{T_1}\sim22.3~\mu$s and $\overline{T_2}\sim4.0~\mu$s. Here, the time-dependent participation entropies can be calculated by the output density matrix $\hat{\rho}(t)$,
corresponding to $p_{i}(t) = \bra{i}\hat{\rho}(t)\ket{i}$, with $|i\rangle$ being the computational basis in $\mathcal{N}$-dimension Hilbert space. Furthermore, post-selection is also employed to mitigate the effect of energy relaxation in the experimental data processing. Taking the above factors into account, the numerical results match with the experimental data quite well, showing the characteristics of phase transition despite the inevitable decoherence.

\section{first-order participation entropy}

\begin{figure*}[ht]
	\centering
	\includegraphics[width=1\linewidth]{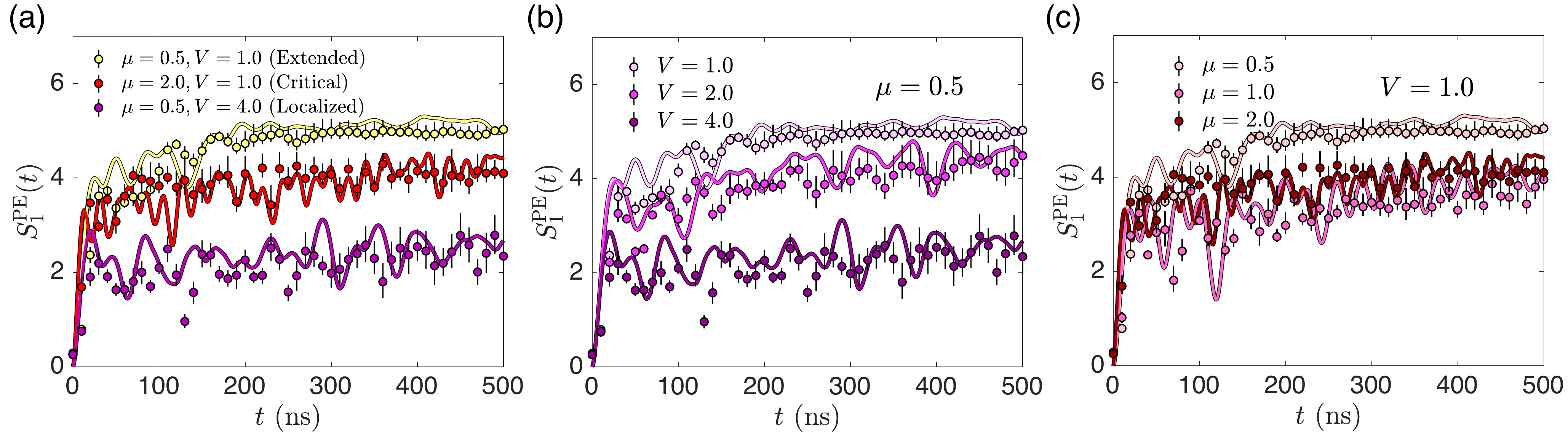}\\
	\caption{(a) The dynamics of first-order participation entropy for the system quenched into the three phases. The yellow, red, and purple points are experimental data for the extended ($\mu=0.5, V=1.0$), critical ($\mu=2.0, V=1.0$), and localized ($\mu=0.5, V=4.0$) phases, respectively. Lines with the same color are numerical simulation under the same parameters with decoherence taken into account. (b) The dynamics of first-order participation entropy with fixed $\mu=0.5$ and $V=1,2$ and $4$ (extended to localized transition). (c) The dynamics of first-order participation entropy with fixed $V=1$ and varying $\mu=0.5, 1$ and $2$ (extended to critical transition).}
    \label{supp_pe1_t}
\end{figure*}

\begin{figure*}[ht]
	\centering
	\includegraphics[width=1\linewidth]{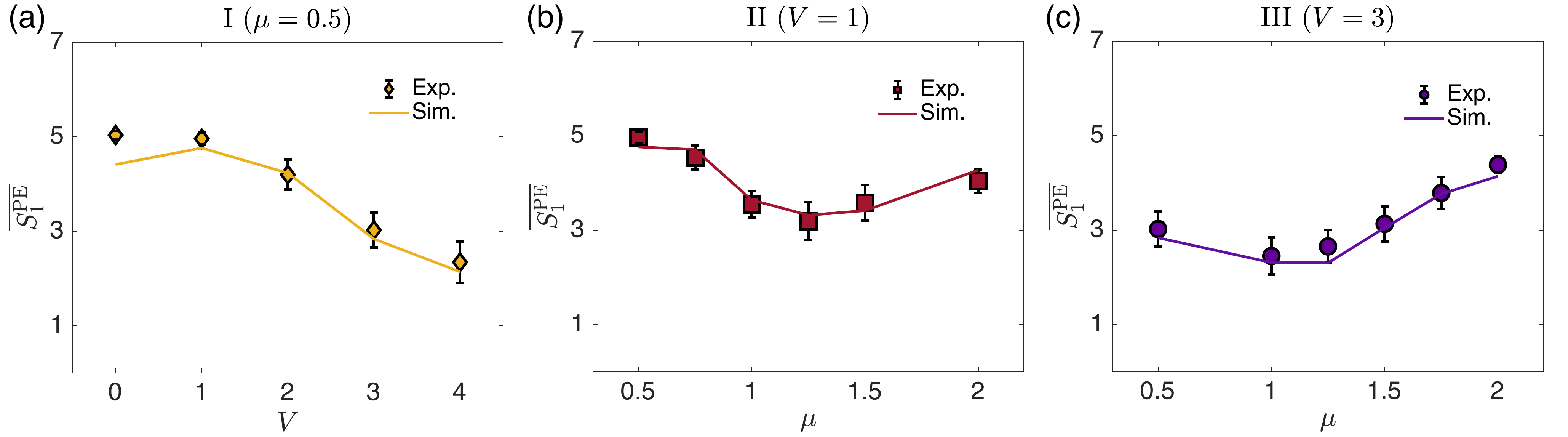}\\
	\caption{Comparisons between experimental data and numerical simulation (a) with fixed $\mu=0.5$ and varying $V$ from extended to localized transition, (b) with fixed $V=1$ and varying $\mu$ from extended to critical transition, and (c) with fixed $V=3$ and varying $\mu$ from localized to critical transition. Points with statistical error bars are experimental data, and solid lines are numerical simulation using the Hamiltonian~\ref{eq_GAAH} in the main text.}
    \label{supp_pe1_ave}
\end{figure*}

In the limit $q \to  1$, we can obtain the dynamical first-order participation entropy:
\begin{equation}
    S^{\text{PE}}_1(t) = \lim\limits_{q \to 1}\frac{1}{1-q}\log{\sum\limits_{i}^{\mathcal{N}} p_{i}(t)^q} = -\sum\limits_{i}^{\mathcal{N}}p_{i}(t)\log{p_{i}(t)}.
\end{equation}
The dynamical behavior of first-order participation entropy is very similar to the second-order participation entropy we discussed above, but with a different $q$-dependent fractal dimension $D_{q=1}$ in terms of scaling behavior.
Here, we also present the dynamics of the first-order participation entropy in Fig.~\ref{supp_pe1_t}, and the averaged late-time first-order participation entropy for different parameters $\mu$ and $V$ in Fig.~\ref{supp_pe1_ave}. Comparing with Fig.~\ref{supp_pe1_t} and Fig.~\ref{fig3} in the main text, we can find that the growth trend over time of first-order participation entropy is the same as that of the second-order, but with slightly higher values than those of the second-order. Thus, the averaged late-time first-order participation entropy can also be used to characterize phase transitions as the second-order, as shown in Fig.~\ref{supp_pe1_ave}.

\section{scaling behavior of participation entropy}
For many-body states, the participation entropy $S^{\text{PE}}_q$ can characterize the localization in the Hilbert space. In the thermodynamic limit $\mathcal{N} \to \infty$, $S^{\text{PE}}_q$ is (\romannumeral1) $\log{\mathcal{N}}$ for a perfectly delocalized state, (\romannumeral2) a const for a localized state, (\romannumeral3) $D_q\log{\mathcal{N}}$ (where $D_q<1$) for a state with a fractal dimension $D_q$, which occurs at localization transition or in the critical phase \cite{Chang1997,Luitz2014}.

\begin{figure}[htbp]
	\centering
	\includegraphics[width=1\linewidth]{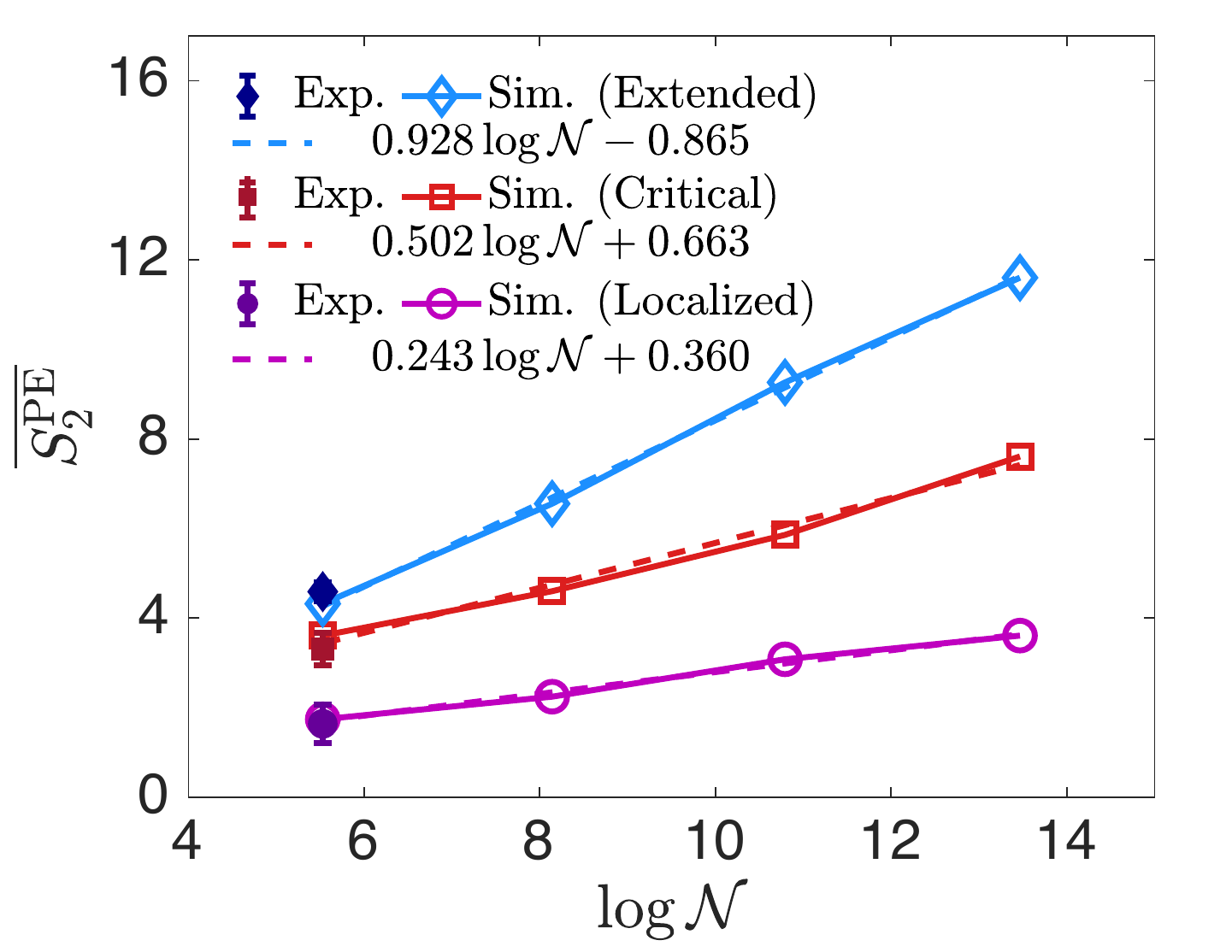}\\
	\caption{The averaged late-time participation entropy as a function of $\log{\mathcal{N}}$. We choose $\mu=0.5, V=1.0$ for the extended phase, $\mu=2.0, V=1.0$ for the critical phase, and $\mu=1.0, V=3.0$ for the localized phase. The results of the fit are drawn with dashed lines.}
    \label{supp_scaling}
\end{figure}

To see this, we display the scaling behavior of the averaged late-time participation entropy $\overline{S^{\text{PE}}_2}$ in Fig.~\ref{supp_scaling}. Since the numerical simulation match with the experimental data quite well, here we use numerical results of larger system sizes to fit with the form $\overline{S^{\text{PE}}_2}=a_2\log(\mathcal{N})+b_2$. We choose three pairs of parameters $(\mu=0.5, V=1.0)$, $(\mu=2.0, V=1.0)$  and $(\mu=1.0, V=3.0)$ for the extended, critical, and localized phases, respectively. In the extended phase, $a_2\approx 0.928$, which is very close to one for a perfectly delocalized state. In the critical phase, multifractal behavior of many-body states is revealed by the fractal dimension $a_2\approx 0.502$. Besides, in the localized phase, $a_2\approx 0.243$ is much smaller than those in the extended and critical phases.

\end{document}